\documentclass[twocolumn,preprintnumbers,amsmath,nofootinbib,amssymb]{revtex4-1}

\usepackage{amsmath,amssymb,epsfig,graphicx}
\usepackage{nicefrac}

\usepackage{dcolumn}

\def\bea{\begin{eqnarray}}
\def\eea{\end{eqnarray}}
\newcommand{\be}{\begin{eqnarray}}
\newcommand{\ee}{\end{eqnarray}}
\newcommand{\ds}{\displaystyle}

\newcommand{\cA}{\mathcal{A}}

\newcommand{\cC}{\mathcal{C}}

\newcommand{\cH}{\mathcal{H}}

\newcommand{\cJ}{\mathcal{J}}

\newcommand{\cM}{\mathcal{M}}
\newcommand{\cN}{\mathcal{N}}
\newcommand{\cO}{\mathcal{O}}

\newcommand{\cS}{\mathcal{S}}
\newcommand{\cT}{\mathcal{T}}

\newcommand{\cV}{\mathcal{V}}

\newcommand{\Z}{\mathbb{Z}}

\newcommand{\R}{\mathbb{R}}

\newcommand{\m}{\mathfrak{m}}
\newcommand{\n}{\mathfrak{n}}
\newcommand{\nn}{\nonumber}
\newcommand{\eg}{{\it e.g.}}

\newcommand{\ie}{{\it i.e.}}

\newcommand{\tell}{{\tilde \ell}}

\begin{document}

\preprint{}

\title{
Star shaped quivers with flux
}

\author{
Shlomo S. Razamat$^1$ and Brian  Willett$^2$
}

\

\affiliation{
{\it $^1$Department of Physics, Technion, Haifa 32000, Israel\\ 
$^2$ Kavli Institute for Theoretical Physics\\
	University of California, Santa Barbara, CA 93106, USA}\\
}

\date{\today}

\begin{abstract}
	We study the compactification of the 6d $\cN=(2,0)$ SCFT on the product of a Riemann surface with flux and a circle.  On the one hand, this can be understood by first reducing on the Riemann surface, giving rise to 4d $\cN=1$ and $\cN=2$ class ${\cal S}$ theories, which we then reduce on $S^1$ to get 3d $\cN=2$ and $\cN=4$ class ${\cal S}$ theories.  On the other hand, we may first compactify on $S^1$ to get the 5d $\cN=2$ Yang-Mills theory.  By studying its reduction on a Riemann surface, we obtain a mirror dual description of 3d class ${\cal S}$ theories, generalizing the star-shaped quiver theories of Benini, Tachikawa, and Xie.  We comment on some global properties of the gauge group in these reductions, and test the dualities by computing various supersymmetric partition functions.  
\end{abstract}


\maketitle

\section{Introduction}

Studying QFTs on compact spaces often leads to insights into complicated dynamics of lower dimensional theories. For example, many dualities between lower dimensional SCFTs can be deduced from dualities connecting higher dimensional theories. A particular example is understanding ${\cal N}=2$ dualities in 3d starting from 4d ${\cal N}=1$ dual theories compactified on a circle \cite{Niarchos:2012ah,Dolan:2011rp,Gadde:2011ia,Aharony:2013dha,Aharony:2013kma}. Another example is the many insights derived in recent years, following the seminal paper \cite{Gaiotto:2009we}, about strong coupling dynamics of 4d ${\cal N}\geq 1$ by understanding them as compactifications of 6d SCFTs on Riemann surfaces. This has lead to improved understanding of dualities and the emergence of symmetry in many examples of 4d SCFTs (see, for example \cite{Benini:2009mz,Bah:2012dg,Gaiotto:2015usa,Kim:2017toz,Razamat:2018gro,Razamat:2019ukg} and references therein). Importantly, the 6d SCFTs here do not have at the moment a useful description in terms of fields and Lagrangians, see \cite{Heckman:2018jxk} for a nice review.

When one considers compactifications of higher dimensional quantum field theory, the resulting lower dimensional model is typically not given just by the KK reduction of the higher dimensional fields with the same types of interactions.  One can understand the problem as follows. In such a setup there are two limits involved: first, we have the computation of the path integral, and second, we have a geometric parameter, the size of the compact part of the geometry, which we take to be small. These two limits need not commute. A concrete example of this is that of taking 4d theories on a circle: the fact that some of the 4d symmetries are anomalous leads to novel interaction terms in the effective 3d theory, which explicitly break the anomalous symmetry \cite{Aharony:2013dha,Aharony:2013kma}. Compactifying 3d  theories to 2d leads to many complications of this sort \cite{Aharony:2017adm}, and similarly for 4d models reduced to 2d \cite{Gadde:2015wta,Dedushenko:2017osi}, and it is fair to say that such reductions are not understood well enough.

In this paper we will discuss some aspects of compactifications of 6d SCFTs down to 3d. We will not consider the compactifications on a generic 3d manifold, as was done for example in \cite{Dimofte:2011ju,Dimofte:2011py,Chung:2014qpa}, but rather on a geometry which has the structure of (punctured) Riemann surface times a circle. There are two ways to view such a compactification. We can either first try to understand the reduction on a circle down to 5d and then a subsequent compactification to 3d, or first compactify to 4d and then to 3d.  
The former way has the advantage that, although the 6d theories are not given in terms of Lagrangians, often when compactified on a circle (possibly with holonomies for various symmetries) they possess an effective 5d description in terms of fields. This 5d description then can be directly used to understand the further compactification down to 3d. This will still be a non trivial task, following the comments we made above, however, we will be able to partially fix the 3d field content and action in terms of the 5d Lagrangian and the flavor symmetry background on the Riemann surface. 

 A useful set of tools in the analysis of dimensional reduction in supersymmetric field theories have been the supersymmetric partition functions on product manifolds.  In this case, we can understand the reduction from 5d to 3d using the $S^3_b \times \Sigma_g$ partition function, studied in \cite{Crichigno:2018adf}, which can be interpreted as the $S^3_b$ partition function of the dimensionally reduced theory.  This partition function is closely related to the twisted index studied in lower dimensions \cite{Nekrasov:2014xaa,Gukov:2015sna,Benini:2015noa,Closset:2016arn}, which has been used to study similar reductions from 4d to 2d in \cite{Gadde:2015wta}.

On the other hand, the approach where we first compactify to 4d often leads to theories which are rather complicated  and typically do not have known Lagrangian descriptions. This makes it harder to understand the further compactifications to 3d. In some well-behaved cases, we can nevertheless study this reduction and the resulting 3d models.  We will then discuss how the two different orders of compactification are related to each other. In particular, this leads to two different dual descriptions of the 3d SCFT obtained in the reduction. 

Although we expect this procedure to apply to more general 6d models, we will concretely discuss compactifications of $A_{N-1}$ $(2,0)$ SCFT and give explicit details for the $A_1$ case, as here the 4d intermediate step is particularly simple. For the cases preserving ${\cal N}=4$ supersymmetry in 3d the dual descriptions are the mirror dualities of \cite{Benini:2010uu}, which take the form of a ``star-shaped quiver''  with a central $SU(N)/\Z_N$ node.  More generally, we may compactify with a flux, $\n$, for the $Sp(1)_F$ 6d flavor symmetry on the Riemann surface, which leads to an $\cN=2$ model in 3d, and here we find simple dual descriptions with a number of adjoint chiral multiplets for the central node which is linear in the flux, $\n$.

The paper is organized as follows. In Section \ref{sec:5dgeneralreduction} we discuss  reductions on a Riemann surfaces with flux of general ${\cal N}=1$ 5d SCFTs and in particular the maximally supersymmetric SYM, which is a circle reduction of the 6d $(2,0)$ theory.
In Section \ref{sec:4dreduction} we then analyze the reductions of 4d ${\cal N}=1$ class ${\cal S}$ theories \cite{Gaiotto:2009we,Gaiotto:2009hg}, which are  obtained by compactifying the $(2,0)$ theory on a Riemann surface with flux on a circle.  Next, the two orders of the reduction are compared in Section \ref{sec:duality} and the ensuing dualities are discussed. In Section \ref{sec:pfchecks} we discuss technical checks of the dualities using various supersymmetric partition functions. We comment on our results and possible generalizations in \ref{sec:disc}.
Appendix \ref{app:tsun} collects several useful properties of the $T[SU(N)]$ models.

\section{Reduction of 5d $\cN=1$ gauge theories on $\Sigma_g$}
\label{sec:5dgeneralreduction}

In this section we describe some general aspects of the reduction of a 5d $\cN=1$ gauge theory on a Riemann surface, $\Sigma_g$, which in general gives rise to a 3d $\cN=2$ theory.  We start by decomposing the fields into modes on $\Sigma_g$ and analyzing the resulting spectrum of fields in 3d.  We then analyze the same problem from the perspective of the $S^3_b \times \Sigma_g$ partition function, computed in \cite{Crichigno:2018adf}, and point out some new features that arise in this analysis.  After this general analysis, we focus on the case of the 5d $\cN=2$ SYM theory, which will be our main example.

\subsection{Modes on $\Sigma_g$}
\label{sec:gf}

A general 5d $\cN=1$ gauge theory has an $Sp(1)_R$ R-symmetry and gauge and flavor symmetries, $G$ and $G_F$.  The matter content consists of a vector multiplet, $V$, in the adjoint representation of $G$, and hypermultiplets, $H_i$, which come in a pseudo-real representation, $R_i \otimes S_i$, of $G \times G_F$.  The field content and representations of these multiplets are as follows:

\

\begin{centering}
	
	\begin{tabular}{|c|c|c|c|c|c|}
		\hline
		Multiplet & Field & $SO(5)_E$ & $Sp(1)_R$ & $G$ & $G_F$ \\
		\hline
		$V$ & $\sigma$ & $1$ & $1$ & Adj & $1$ \\
		& $\Lambda$ & $4$ & $2$ & Adj & $1$ \\
		& $A_\mu$ & $5$ & $1$ & Adj & $1$ \\	 
		\hline
		$H_i$ & $Q_i$ & $1$ & $2$ & $R_i$ & $S_i$ \\
		& $\Psi_i$ & $4$ & $1$ & $R_i$ & $S_i$ \\
		\hline
	\end{tabular}
	
\end{centering}

\

\noindent Here we impose reality conditions on $\Lambda$, $Q_i$, and $\Psi_i$, which each live in a tensor product of two pseudo-real representations. The group $SO(5)_E$ is the group of rotations of the 5d Euclidean space.

We consider the theory on the spacetime $\R^3 \times \Sigma_g$, and to preserve supersymmetry we perform a partial topological twist along $\Sigma_g$.  This consists of introducing a background R-symmetry gauge field in the $U(1)_R$ maximal torus of $Sp(1)_R$, which we identify with the spin connection on $\Sigma_g$.  In addition, we may introduce arbitrary GNO fluxes for $G \times G_F$, which take values in the coweight lattices of these groups, and we denote these fluxes by $(\m,\n)$.  The symmetry which is unbroken in 3d after this twist consists of $U(1)_R$ and the commutant of $G$ and $G_F$ with the fluxes that we turn on.  In principle we should consider the contribution from all  the dynamical gauge fluxes, $\m$, but we will argue that in favorable cases only the sector with $\m=0$ contributes, so for now we will specialize to this case.

Let us start by decomposing these fields into modes on $\Sigma_g$, each of which gives rise to a 3d field.  First we decompose into representations of $SO(3)_E \times U(1)_{\Sigma_g} \subset SO(5)_E$, under which $4 \rightarrow 2_{\pm 1}$ and $5 \rightarrow 3_0 \oplus 1_{\pm 2}$.   Recall that (left/right-handed) fermions on $\Sigma_g$ are sections of the line bundle $\cO(\pm(g-1))$, while $1$-forms are sections of $\cO(\pm 2(g-1))$.  Thus, before the twist, the fields take values in the following sections of $\Sigma_g$:

\

\begin{centering}
	
	\begin{tabular}{|c|c|c|c|c|c|c|}
		\hline
		Multiplet & Field & Bundle & $SO(3)_E$ & $U(1)_R$ & $G$ & $G_F$ \\
		\hline
		$V$ & $\sigma$  & $\cO(0)$ & $1$ & $0$ & Adj & $1$ \\
		& $\Lambda$ & $\cO(\pm (g-1))$ & $2$ & $-1$ & Adj & $1$ \\
		& $A_\mu$ & $\cO(0)$ & $3$ & $0$ & Adj & $1$ \\	 
		& $A_z$ & $\cO(2(g-1))$ & $1$ & $0$ & Adj & $1$ \\	 
		\hline
		$H_i$ & $Q_i$ & $\cO(0)$ & $1$ & $1$ & $R_i$ & $S_i$ \\
		& $\Psi_i$ & $\cO(g-1)$ & $2$ & $0$ & $R_i$ & $S_i$ \\
		\hline
	\end{tabular}
	
\end{centering}

\

Now we consider the effect of the twist.  First we look at the vector multiplet.  Here only the $U(1)_R$ flux has an effect, and after the twist the fields behave as:

\

\begin{centering}
	
	\begin{tabular}{|c|c|c|c|c|c|c|}
		\hline
		Multiplet & Field & Bundle & $SO(3)_E$ & $U(1)_R$ & $G$ & $G_F$ \\
		\hline
		$V$ & $\sigma$  & $\cO(0)$ & $1$ & $0$ & Adj & $1$ \\
		& $\Lambda$ & $\cO(0)$ & $2$ & $-1$ & Adj & $1$ \\
		& $A_\mu$ & $\cO(0)$ & $3$ & $0$ & Adj & $1$ \\	 
		& $A_z$ & $\cO(2(g-1))$ & $1$ & $0$ & Adj & $1$ \\	 
		& $\Lambda_z$ & $\cO(2(g-1))$ & $2$ & $1$ & Adj & $1$ \\
		\hline
	\end{tabular}
	
\end{centering}

\

\noindent We see that $(\sigma,\Lambda,A_\mu)$ contain the fields of a 3d $\cN=2$ vector multiplet.  In principle we obtain such a multiplet for each mode on $\Sigma_g$, however all of the non-holomorphic modes will pair up into long multiplets and decouple from BPS observables, and so we obtain a single 3d $\cN=2$ vector multiplet from the single (constant) zero mode of $\cO(0)$.  In addition, $(A_z,\Lambda_z)$ transform like a $1$-form on $\Sigma_g$, and we expect $g$ holomorphic zero modes for this bundle, which leads to $g$ 3d $\cN=2$ adjoint chiral multiplets. 

Next consider the hypermultiplet, $H_i$.  Recall this sits in a pseudo-real representation, $R_i \otimes S_i$, of $G \times G_F$, and so the weights come in pairs, $(\rho,\omega)$ and  $(-\rho,-\omega)$, and let us consider the contribution from one such pair.  Then these fields get an additional contribution to their flux of $\pm \omega(\n)$.  The resulting field content after the twist is then:

\

\begin{centering}
	
	\begin{tabular}{|c|c|c|c|c|c|c|}
		\hline
		Multiplet & Field & Bundle & $SO(3)_E$ & $U(1)_R$ & $G$ & $G_F$ \\
		\hline
		$H_i$ & $Q_i$ & $\cO(g-1+\omega(\n))$ & $1$ & $1$ & $\rho$ & $\omega$ \\
		& $\Psi_i$ & $\cO(g-1+\omega(\n))$ & $2$ & $0$ & $\rho$ & $\omega$ \\
		& $\tilde{Q}_i$ & $\cO(g-1-\omega(\n))$ & $1$ & $1$ & $-\rho$ & $-\omega$ \\
		& $\tilde{\Psi}_i$ & $\cO(g-1-\omega(\n))$ & $2$ & $0$ & $-\rho$ & $-\omega$ \\
		\hline
	\end{tabular}
	
\end{centering}

\

\noindent Here the number of holomorphic sections of these bundles will depend on the precise metric and gauge connection we choose, however, this number is constrained by the Riemann-Roch index theorem, which states:
\bea \text{dim} H_0(\Sigma_g,\cO(n)) - \text{dim} H_0(\Sigma_g,\cO(2(g-1)-n)) \nn \\ = n + 1-g  \;\;\;\;\;\;\;\;\;\eea
Thus we find some number $N_{\rho,\omega}$ of 3d $\cN=2$ chiral multiplets transforming with weight $(\rho,\omega)$, and $N_{-\rho,-\omega}$ with weight $(-\rho,-\omega)$, subject to:
\be \label{chiredcond} N_{\rho,\omega} - N_{-\rho,-\omega} = \omega(\n) \ee
Importantly, BPS observables, such as supersymmetric partition functions, will depend only on this difference, however the precise field content can not be determined by this analysis.

To summarize the analysis above, we have found that:

\begin{eqnarray} \label{5dvecred}
& \text{5d} \; \cN=1 \;\;\text{vector multiplet} \\
& \Bigg\downarrow \;\;\; \text{compactification  on} \;  \Sigma_g \; \text{with flux} \; \n \nn \\ 
& \text{3d} \; \cN=2  \;\;\text{vector multiplet} \;\;\;\; + \nn \\
& \;\;\;\; g \times \text{3d} \; \cN=2\; \;  \text{adjoint chiral multiplets of R-charge zero} \nn 
\end{eqnarray}

\begin{eqnarray} \nn \label{5dchired}
& \text{5d} \; \cN=1 \;\;\text{hyper multiplet with weight} \; (\rho,\omega) \\
& \Bigg\downarrow  \;\;\; \text{compactification  on} \;  \Sigma_g  \; \text{with flux} \; \n \nonumber \\ 
&  N_{\rho,\omega} \;\times\; \text{3d} 
\; \cN=2  \;\;\text{chirals with weight  $(\rho,\omega)$} \nonumber \\
&+\;N_{-\rho,-\omega} \;\times\; \text{3d} \; \cN=2 \; \text{chirals with weight $(-\rho,-\omega)$} \nonumber \\
& \text{ both with R-charge $1$, and with $N_{\rho,\omega} - N_{-\rho,-\omega} = \omega(\n)$} \nonumber
\end{eqnarray}

We emphasize that the above analysis was performed in the zero gauge flux sector.  We comment below on the possibility of other gauge flux sectors contributing in 3d.

\subsection{The $S^3_b \times \Sigma_g$ partition function}
\label{sec:s3bsgred}

We can gain another perspective on the reduction by studying the $S^3_b \times \Sigma_g$ partition function, computed in \cite{Crichigno:2018adf}.  Let us fix a 5d $\cN=1$ theory with semisimple gauge group $G$, and with matter in a representation $R$ of the gauge and flavor group, $G \times G_F$.\footnote{One could also include a 5d Chern-Simons term, but we will not consider this case here.}  Then, from \cite{Crichigno:2018adf}, the partition function on $S^3_b \times \Sigma_g$ may be written as\footnote{We refer to \cite{Crichigno:2018adf} for more details and conventions.}
\bea &\ds Z_{S^3_b \times \Sigma_g}(\nu)_\n = \frac{1}{|W_G|} \sum_{\m \in \Lambda_G} \nn \\
& \ds \times \int du \; e^{2 \pi i \gamma Tr (u \m) } \prod_{\alpha \in Ad(G)'} s_b(-i Q + \alpha(u))^{-\alpha(\m)+1-g} \nn \\
& \ds \times \prod_{(\rho,\omega) \in R} s_b(\rho(u)+\omega(\nu))^{\rho(\m)+\omega(\n)} H^g \; Z_{inst}(u,\nu)_{\m,\n} \nn  \;.\eea
Here $s_b(x)$ is the double sine function (see {\it e.g.}  \cite{Hama:2011ea}), $u$ and $\m$ parameterize the gauge vector multiplet, determining the real scalar and flux through $\Sigma_g$, respectively, and these parameterize the BPS locus we integrate over after applying localization.  Similarly, $\nu$ and $\n$ describe background vector multiplets coupled to the flavor symmetry, and are parameters of the partition function.  In addition, we have defined $\gamma = -\frac{2 \pi Q}{{g_5}^2}$, where $Q=\frac{b+b^{-1}}{2}$ and $g_5$ is the 5d gauge coupling, and $Ad(G)'$ refers to the non-zero roots of $G$.  The perturbative Hessian, $H$, is given by
\bea \label{Hdef}&\ds H = \det_{ab} \bigg( \gamma K_{ab} + \sum_{(\rho,\omega) \in R} \rho_a \rho_b \frac{s_b'(\rho(u)+\omega(\nu))}{s_b(\rho(u)+\omega(\nu))} \nn \\
& \ds -\sum_{\alpha \in Ad(G)'} \frac{s_b'(-i Q+\alpha(u))}{s_b(-i Q+\alpha(u))} \bigg) \;. \eea
Finally, $Z_{inst}(u,\nu)_{\m,\n}$ refers to the instanton contribution to the partition function, which we do not write explicitly here.

This partition function has a similar structure to the $S^3$ partition function \cite{Kapustin:2009kz,Jafferis:2011ns,Hama:2011ea} of a 3d $\cN=2$ gauge theory, including the expected $1$-loop contributions of the chiral and vector multiplets, expressed through the double sine function, $s_b(x)$.  However, there are two important differences.  First, there is an infinite sum over flux sectors, $\m$, on $\Sigma_g$.  We may tentatively interpret this as implying the system is described by an infinite direct sum of 3d $\cN=2$ theories.  Similar direct sums have appeared in other examples of dimensional reduction of gauge theories, \eg, in \cite{Hwang:2017nop,Aharony:2017adm,Hwang:2018riu}.  In addition, there are extra factors related to the fermion zero modes and instanton contributions, which are not straightforwardly interpreted in terms of the $S^3_b$ partition function of ordinary 3d $\cN=2$ gauge theories, implying this would have to be a more exotic 3d $\cN=2$ theory.

Below, we will focus on a special case where the 3d interpretation is more straightforward.  There is a class of 5d gauge theories which are believed to have a UV completion as a 6d $\cN=(1,0)$ SCFT.  Specifically, there is an emergent circle, whose radius is proportional to the 5d gauge coupling, $\beta \sim g_5$.  In these cases, the partition function above can be reinterpreted as the $S^3_b \times \Sigma_g \times S^1_\beta$ partition function of this 6d theory, or equivalently, as the $S^3_b \times S^1_\beta$ partition function of the 4d theory obtained by compactification on $\Sigma_g$.  If we now consider the limit $\beta \sim g_5 \rightarrow 0$, we may interpret this as the dimensional reduction of the 4d theory, which we expect to give an ordinary 3d theory.

Several things happen in this limit of the partition function above.  First, and most importantly, we expect that the instanton contribution drastically simplifies, and can be essentially ignored.  In fact, we will argue below it contributes an overall factor which can be related to the Cardy behavior of the 4d index as $\beta \rightarrow 0$.

Next note that for $\m \neq 0$, because of the $e^{2 \pi i \gamma Tr (u \m) } $ factor, and since $\gamma \sim \beta^{-1}$ is taken very large, the integrand is rapidly oscillating in at least one direction in the complex $u$ plane, and so by the Riemann-Lebesgue lemma, we expect its contribution to be exponentially small in $\gamma^{-1}$.  In addition, we note that the first term in \eqref{Hdef} is dominant, and so we may approximate,
\be H \approx \gamma^{r_G} \;, \ee
where $r_G$ is the rank of $G$.  More precisely, these two statements only follow provided the integrand is bounded at infinity.  But, as argued in Section $4.2.2$ of \cite{Crichigno:2018adf}, this is true precisely for those 5d theories which have 6d uplifts.

With these assumptions, we may approximate
\bea \label{s3sg1} & \ds Z_{S^3_b \times \Sigma_g}(\nu)_\n \underset{g_5 \rightarrow 0}{\approx} \frac{\gamma^{g \;r_G }}{|W_G|} \int du \prod_{\alpha \in Ad(G)'} s_b(-i Q + \alpha(u))^{1-g} \nn \\
 &\ds \prod_{(\rho,\omega) \in R} s_b(\rho(u)+\omega(\nu))^{\omega(\n)}  \;. \eea
In this form, the partition function looks very similar to the $S^3_b$ partition function of a certain 3d $\cN=2$ gauge theory.  In fact, we claim this theory is precisely the one we were led to by the analysis of the previous subsection.  

To see this, note that the first product in the integrand can be interpreted as the contribution of a 3d $\cN=2$ vector multiplet, along with $g$ adjoint chiral multiplets of R-charge zero, as in \eqref{5dvecred}.  The latter contribute oppositely to the vector, so their total contribution appears as a single factor raised to the power $1-g$.  To be precise, the adjoint chiral multiplets have an additional contribution from the $g r_G$ Cartan elements, which do not depend on the gauge variable, $u$.  Since they also have R-charge zero, strictly speaking their contribution diverges.  However, we may schematically identify this divergence with the $\gamma^{g r_G}$ prefactor, which is also diverging in this limit, and we note the exponent matches the number of Cartan elements.

Next we look at the second product in the integrand in \eqref{s3sg1}.  Using the basic identity,
\be s_b(u) = s_b(-u)^{-1} \;, \ee
we see that we may interpret this as the contribution of $N_{\rho,\omega}$ chiral multiplets of weights  $(\rho,\omega)$ along with $N_{-\rho,-\omega}$ of weight $(-\rho,-\omega)$, provided that
\be N_{\rho,\omega} - N_{-\rho,-\omega} = \omega(\n) \;, \ee
precisely as in \eqref{5dchired}.  

Thus, under the assumptions above which led us to argue the partition function truncates to the zero flux sector, we see the analysis of the previous subsection and that of the $S^3_b \times \Sigma_g$ partition function lead to the same result.

\subsubsection*{Cardy scaling}

From the results of \cite{DiPietro:2014bca}, we expect the 3d reduction of the 4d index to behave in the $\beta \rightarrow 0$ limit as
\be \label{cardy} \lim_{\beta \rightarrow 0} {\cal I}(p,q,\mu_i) = \exp \bigg( -\frac{\pi}{6\beta} \bigg( Q \cA^R + \cA^\alpha \nu_\alpha \bigg) \bigg) Z_{S^3_b}(
\nu_i) \nn \;, \ee 
where $\cA^R$ and $\cA^\alpha$ are the linear anomalies of the R-symmetry and flavor symmetries, respectively.  Thus, while above we have found the expected finite piece, we did not observe the divergent ``Cardy scaling.''  We expect this contribution will arise from the instanton contribution we have  so far ignored.  We will see this more explicitly when we consider the 5d $\cN=2$ theory next.

\subsection{Reduction of the 5d $\cN=2$ SYM theory}
\label{sec:5dN2reduction}

Our main example in this paper will be the 5d $\cN=2$ Yang-Mills theory with gauge group $G$, which we will usually take to be $SU(N)$.  This has an $Sp(2)_R$ symmetry, but in the $\cN=1$ language used above, this decomposes to $Sp(1)_R \times Sp(1)_F$, with a single 5d $\cN=1$ vector multiplet and a 5d $\cN=1$ hypermultiplet in the representation $(\text{Adj},2)$ of $G \times Sp(1)_F$.

Our main interest in this example is due to the fact that it admits a UV completion as the 6d $\cN=(2,0)$ SCFT compactified on a circle.  Then the 3d reduction of this 5d theory on $\Sigma_g$ may be alternatively described in terms of first compactifying the 6d SCFT on a Riemann surface with flux $\n$ for the $Sp(1)_F$ flavor symmetry, obtaining in  general a 4d $\cN=1$ class ${\cal S}$ theory \cite{Benini:2009mz,Bah:2011je,Bah:2012dg} (see also \cite{Agarwal:2015vla,Nardoni:2016ffl,Fazzi:2016eec}),\footnote{In the notation of \cite{Bah:2011je}, $\n=\frac{p-q}{2}$, as discussed below.}  and then dimensionally reducing this on a circle.  Comparing the theories obtained by these two methods can then lead to non-trivial three dimensional dualities  We will return to the 4d class ${\cal S}$ theories and their reduction in the next section.

\subsubsection*{Reduction on $\Sigma_g$}

\begin{table}[t]
	
	\begin{centering}
		
		\begin{tabular}{|c|c|c|c|c|}
			\hline
			Multiplet & Number & $U(1)_R$ & $G$ & $U(1)_F$ \\
			\hline
			$V$ & $1$ & - & Adj & $0$ \\
			$\Omega_a$ & $g$ & $0$ & Adj & $0$ \\
			$\Psi_b$ & $\ell$ & $1$ & Adj & $1$ \\
			$\tilde{\Psi}_c$ & $\tell$ & $1$ & Adj & $-1$ \\
			\hline
		\end{tabular}
		
	\end{centering}
	
	\caption{Matter content of 3d reduction of 5d $\cN=2$ SYM.  Here $V$ denotes a 3d $\cN=2$ vector multiplet and the remaining multiplets are chiral multiplets, where the $U(1)_R$ charge refers to that of the scalar component.}
	\label{n2redtab}		
\end{table}

When compactifying the 5d $\cN=2$ theory on $\Sigma_g$, we may also include a flux, $\n \in \Z$, for the $Sp(1)_F$ flavor symmetry, and so we expect a family of theories labeled by $\n$ and $g$.  Let us first analyze this reduction in terms of the modes on $\Sigma_g$, as in Section \ref{sec:gf}.  We find the field content in 3d shown in Table \ref{n2redtab}.  Here the numbers, $\ell$ and $\tell$, of adjoint chiral multiplets, $\Psi$ and $\tilde{\Psi}$, cannot be determined individually, but satisfy
\be \ell - \tell = \n \;. \ee

Let us first consider an important special case, which is $\n=g-1$.  This can be interpreted as performing the topological twist on $\Sigma_g$ using the R-symmetry
\be R_+ = R + F \;.\ee
This is the same twist used to define 4d $\cN=2$ class ${\cal S}$ theories, and so we expect the 3d theories obtained here to be equivalent to their dimensional reduction.  In this case, it is natural to take the background gauge field equal to the spin connection on $\Sigma_g$, and in this case the fields $\Psi$ and $\tilde{\Psi}$ are sections of $\cO(2(g-1))$ and $\cO(0)$, respectively, which can be identified with $1$-forms and scalars.  Then we are justified in taking
\be \ell = g, \;\;\; \tell = 1  \;. \ee
Then we see the fields can be organized into the field content of 3d $\cN=4$ theory, namely,
\bea & \ds (V, \; \tilde{\Psi}_1) & \;\;\; \rightarrow \;\;\; \text{3d $\cN=4$ vector multiplet,}  \\
& \ds (\Omega_a, \; \Psi_a) & \;\;\; \rightarrow \;\;\; \text{$g \; \times$ 3d $\cN=4$ adjoint hypermultiplets.} \nn \eea

Let us check if the symmetries act as expected for a 3d $\cN=4$ theory, whose symmetry group is $SL(2,\R)_{rot} \times SU(2)_H \times SU(2)_C$.
Decomposing the R-symmetry under the $U(1)_H \times U(1)_C$ maximal torus, and allowing also a $U(1)_f$ flavor symmetry to act on the hypermultiplet, we see the expected behavior is:

\

\begin{centering}
	
	\begin{tabular}{|c|c|c|c|c|c|c|c|}
		\hline
		 $\cN=4$ & $\cN=2$ & Field & $SL(2,\R)_{rot}$ & $U(1)_H$ & $U(1)_C$ & $U(1)_f$ \\
		\hline
		$\text{Vector}$ & $V$ & $\sigma$  & $1$ & $0$ & $0$ & $0$\\
		& & $\Lambda$ &  $2$ & $-1$ & $-1$ & $0$ \\
		& & $A_\mu$ & $3$ & $0$ & $0$ & $0$ \\	 
		& $\Phi$ & $\phi$ &  $1$ & $0$ & $2$ & $0$ \\	 
		& & $\psi_\phi$ & $2$ & $-1$ & $1$ & $0$ \\
		\hline
		$\text{Hyper}$ & $Q$ & $q$ & $1$ & $1$ & $0$ & $1$ \\
		& & $\psi$ & $2$ & $0$ & $-1$ & $1$ \\
		& $\tilde{Q}$ & $\tilde{q}$ & $1$ & $1$ & $0$ & $-1$ \\
		& & $\tilde{\psi}$ & $2$ & $0$ & $-1$ & $-1$ \\
		\hline
	\end{tabular}
	
\end{centering}

\

\noindent On the other hand, we can consider the $U(1)_R \times U(1)_F$ symmetry acting on the fields obtained from 5d.  Here we include also the charges under a $U(1)_f$ flavor symmetry, which sits inside the $U(g)$ symmetry acting on the fields $(\Omega_a,\Psi_a)$, which is a hidden symmetry from the 5d point of view:

\

\begin{centering}
	
	\begin{tabular}{|c|c|c|c|c|c|c|}
		\hline
		Multiplet & Field & $SO(3)_E$ & $U(1)_R$ & $U(1)_F$ & $U(1)_f$ \\
		\hline
		$V$ & $\sigma$  & $1$ & $0$ & $0$ & $0$ \\
		& $\Lambda$ &  $2$ & $-1$ & $0$ & $0$ \\
		& $A_\mu$ & $3$ & $0$ & $0$  & $0$\\
		\hline
		$\Omega_a$ & $A_z$ &  $1$ & $0$ & $0$ & $1$ \\	  
		& $\Lambda_z$ & $2$ & $-1$ & $0$ & $1$ \\
		\hline
		$\Psi_b$ & $Q_z$ & $1$ & $1$ & $1$ & $-1$ \\
		& $\Psi_z$ & $2$ & $0$ & $1$ & $-1$ \\
		\hline
		$\tilde{\Psi}_1$ & $\tilde{Q}$ & $1$ & $1$ & $-1$ & $0$ \\
		& $\tilde{\Psi}$ & $2$ & $0$ & $-1$ & $0$ \\
		\hline
	\end{tabular}
	
\end{centering}

\

Then we see the charges agree provided we identify
\be \label{hidmix} H = \begin{cases} R+F & g=0 \\ R+F+ f & g>0 \end{cases}, \;\;\;\; C = R-F \;. \ee
Interestingly, to identify these symmetries, we see that for $g>0$, we must admix a flavor symmetry which is hidden from the 5d point of view.  Note that this is very reminiscent of Gaiotto-Witten ``bad'' theories having hidden IR symmetry \cite{Gaiotto:2008ak}.

As mentioned above, we may also interpret this theory as the $S^1$ reduction of a 4d $\cN=2$ class ${\cal S}$ theory, obtained by compactifying the 6d $\cN=(2,0)$ theory on a Riemann surface.  The 3d reduction of the class ${\cal S}$ theory associated to a Riemann surface has a dual description, found in \cite{Benini:2010uu}, as a so-called ``star-shaped quiver.''  In the present case, with no flavor punctures, this is a 3d $\cN=4$ theory with $g$ adjoint hypermultiplets.  This is precisely the description we have found above, providing an alternative derivation of their result.

Note that when we twist by the $U(1)_{R_+} \subset Sp(1)_R \times Sp(1)_F \subset Sp(2)_R$ symmetry in 5d, the commutant is an $SU(2)$ subgroup, with Cartan $R_-=R-F$, which, using \eqref{hidmix}, we can identify with the $SU(2)_C$ symmetry.  On the other hand, if we consider this 5d theory as a 6d theory compactified on a circle, this $SU(2)$ commutant becomes the $SU(2)_R$ symmetry of the resulting 4d $\cN=2$ class ${\cal S}$ theory.  In the usual convention, this $SU(2)_R$ symmetry becomes, upon dimensional reduction, the $SU(2)_H$ symmetry of the resulting 3d $\cN=4$ class ${\cal S}$ theory.  The fact that it acts as an $SU(2)_C$ symmetry above reflects the fact that this description should  be considered a ``mirror dual'' of the 3d $\cN=4$ class ${\cal S}$ theory.

For future reference, we will find it useful to define $\cN=2^*$ $U(1)_r$ symmetry and $U(1)_t$ flavor symmetries as follows:
	\bea \label{rtdef}  & \ds r = \frac{1}{2}(H+C) = \begin{cases} R & g=0 \\ R+\frac{f}{2} & g>0 \end{cases} , \nn \\
	& \ds {T} = \frac {1}{2}(C-H) = \begin{cases} - F & g=0 \\ - F - \frac{f}{2} & g>0 \end{cases}  .  \eea
Here we have defined the $U(1)_t$ flavor symmetry with a sign relative to the usual convention.  This is in anticipation of comparison to the dimensional reduction  of 4d models, where, given the previous paragraph, we expect the usual $U(1)_t$ symmetry to map to the one with a flip of sign defined above.

The above argument can be generalized for arbitrary flux, $\n$, for the $U(1)_F$ symmetry, which will in general lead to a 3d $\cN=2$ theory. These theories may alternatively be obtained by dimensional reduction of 4d $\cN=1$ class ${\cal S}$ theories associated to compactification  on a Riemann surface with flux \cite{Benini:2009mz,Bah:2011je,Bah:2012dg}, which we describe in more detail in the next section.  For general $\n$, the matter content can be summarized by Table \ref{n2redtab} above, and we note the charges are compatible with the superpotential
\be \label{superpot} W = \sum \Omega_a \Psi_b \tilde{\Psi}_c  \;,\ee
where the sum is over any subset of the allowed values of the indices.  For example, in the $\cN=4$ case, where $\n=g-1$, the superpotential in \eqref{superpot} may be taken as
\be W = \sum_{a=1}^g \Omega_a \tilde{\Psi}_1 \Psi_a \;,\ee
which is the appropriate 3d $\cN=4$ superpotential.  A similar statement holds for the case $\n=-(g-1)$, with the roles of $\Psi$ and $\tilde{\Psi}$ exchanged.

Another interesting example is the case $\n=0$.  Then $\ell=\tell$, so the adjoint chiral multiplets come in $\ell$ pairs, and, although we can't fix them individually, we can see that each such pair forms a doublet of the $SU(2)_F$ symmetry, which remains unbroken in this case. The 4d parents of these theories are the so-called ``Sicilian'' 4d $\cN=1$ theories \cite{Benini:2009mz}.

To summarize, we find that the 3d theory corresponding to the compactification of the 6d $\cN=(2,0)$ $A_{N-1}$ theory on a Riemann surface of genus $g$ and with flux $\n$ for the $U(1)_F \subset SU(2)_F$ flavor symmetry is described by
\be \label{3dssqs0} & \ds \text{ $\frak{su}(N)$ gauge theory with $g, \; \ell$, and $\tell$} \nn \\
& \ds \text{adjoint chirals of $U(1)_F$ charge $0,\;1$ and $-1$}  \nn \\
& \ds \text{where $\ell-\tell=\n$.} \ee
In Section \ref{sec:global} below we will see that the global form of the gauge group is naturally taken to be $SU(N)/\Z_N$.
	
\subsubsection*{$S^3_b \times \Sigma_g$ partition function}

Let us now briefly reconsider the above analysis using the $S^3_b \times \Sigma_g$ partition function.  In this case the perturbative partition function is given by
\bea & \ds Z_{S^3_b \times \Sigma_g}^{\cN=2,pert}(\nu)_\n = \frac{1}{|W_G|} \sum_{\m \in \Lambda_G} \nn \\
& \ds \times  \int du  e^{2 \pi i \gamma Tr (u \m) } \prod_{\alpha \in Ad(G)} s_b(\alpha(u)+\nu)^{\alpha(\m)+\n} \nn \\
& \ds \times \prod_{\alpha \in Ad(G)'}  s_b(-i Q + \alpha(u))^{-\alpha(\m)+1-g} H^g \;. \eea
Here $\nu$ and $\n$ are the mass and flux, respectively, for the $SU(2)_F$.  Also, $\gamma = -\frac{2 \pi Q}{{g_5}^2}$, where $Q=\frac{b+b^{-1}}{2}$.  

Now we consider the limit $g_5 \rightarrow 0$, or equivalently, $\gamma \rightarrow \infty$.  As above, in this limit we expect to be justified in considering only the perturbative contribution and the $\m=0$ term in  the sum over fluxes, and we find
\bea \label{s3sg12} & \ds Z_{S^3_b \times \Sigma_g}^{\cN=2,pert}(\nu)_\n \underset{g_5 \rightarrow 0}{\approx} \frac{\gamma^{g \;r_G }}{|W_G|} \int du \prod_{\alpha \in Ad(G)} s_b(\alpha(u)+\nu)^{\n}  \nn \\
& \ds \prod_{\alpha \in Ad(G)'} s_b(-i Q + \alpha(u))^{1-g} \;. \eea

Let us first consider the case $\n=g-1$ (${\n}=-(g-1)$ is analogous), corresponding to reduction preserving 3d $\cN=4$ supersymmetry.  Then we may write
\bea \label{s3sg2} & \ds Z_{S^3_b \times \Sigma_g}^{\cN=2,pert}(\nu)_{\n=g-1} \underset{\gamma \rightarrow \infty}{\approx}  \\
& \ds \frac{\gamma^{g r_G}s_b(\nu)^{r_G(g-1)}}{|W_G|} \int du \prod_{\alpha \in Ad(G)'} \bigg( \frac{s_b(-i Q + \alpha(u))}{s_b(\alpha(u)+\nu)}\bigg)^{1-g} \nn \;. \eea
Let us compare this to the $S^3_b$ partition function of the star-shaped quiver of \cite{Benini:2009mz}.  Here we use the standard $U(1)_r$ and $U(1)_t$ symmetries, as in \eqref{rtdef}, as well as the $U(g)$ flavor symmetry, and denote their fugacities by $\tau$ and $\mu_i$, $i=1,...,g$, respectively, giving,
\bea\label{ssq} &\ds Z_{S^3_b}^{SSQ}(\mu_i,\tau) = \frac{1}{|W_G|} \int du \prod_{\alpha \in Ad(G)'}  s_b(-iQ+\alpha(u))  \;\;\;\;\;\;  \\
& \ds \times \prod_{\alpha \in Ad(G)}\bigg( s_b( \tau+\alpha(u)) \prod_{i=1}^g s_b(\frac{iQ}{2} - \frac{\tau}{2}\pm \mu_i+\alpha(u)) \bigg) \;, \nn \eea
where the first two factors in the integrand come from the 3d $\cN=4$ vector multiplet, and the remaining factors come from the adjoint hypers.  This $U(g)$ symmetry is accidental from the point of view of the 5d theory, and so the limit of the 4d index does not give us access to the full symmetry, but places us at a particular point in the space of real mass parameters.  In fact, following the analysis that led to \eqref{hidmix}, one finds that we should identify
\be \tau = -\nu, \;\;\;\;  \mu_i = \frac{-i Q + \nu}{2} \;. \ee
Then we find:
\bea \label{ssq2} & \ds Z_{S^3_b}^{SSQ}  = s_b(\nu)^{r_G(g-1)} s_b(i Q)^{g r_G} \nn \\
& \ds \times   \frac{1}{|W_G|} \int du
\prod_{\alpha \in Ad(G)'} \bigg( \frac{s_b(-iQ+\alpha(u))}{s_b(\nu+\alpha(u))} \bigg)^{1-g} \;. \eea
This expression is somewhat formal for $g>0$, as $s_b(x)$ has a  simple pole at $x=i Q$.  However, comparing to \eqref{s3sg2}, we see the finite pieces precisely agree, and the divergences also formally match if we identify $s_b(i Q) \sim \gamma \rightarrow \infty$.  A similar analysis can be carried out for more general flux, $\n$, and we find the $S^3_b$ partition function of the 3d $\cN=2$ theory described above.

\subsubsection*{Cardy scaling and the Schur limit}

As mentioned above, we expect the instantons, which we have so far ignored, will contribute the expected Cardy scaling of the 4d index as $\beta \rightarrow 0$, as in \eqref{cardy}.  First, to see what result we expect, the anomaly polynomial of the general 4d $\cN=1$ theory above was computed in \cite{Bah:2011je}, and in particular,
\be \text{Tr} \; {\cal R} = (g-1) r_G  (1+z \epsilon) \;,\ee
where $z=\frac{\n}{(g-1)}$ and $\epsilon$ is the mixing parameter of the R-symmetry with the $U(1)_t$ flavor symmetry.  After adapting to our  notation, this leads to a predicted Cardy scaling of
\be \label{4dscal}  \exp \bigg( -\frac{\pi \gamma}{12} (N-1) \bigg( (b +b^{-1})(g-1) -2 i \nu \n \bigg) \bigg) \nn \;.\ee

The $S^3_b \times \Sigma_g$ partition function of the $\cN=2$ SYM theory admits a special limit with enhanced supersymmetry, called the ``Schur limit'' in \cite{Crichigno:2018adf} due to its relation to the Schur limit of the 4d index \cite{Gadde:2011ik}.  This corresponds to setting
\be \nu = \frac{i}{2} (b-b^{-1}) \;. \ee
In this limit the partition function greatly simplifies, and one can compute the instanton contribution is given by
\be Z_{inst}(\nu)_\n = \eta(z)^{(N-1)(g-1+\n)} \eta(\bar{z})^{(N-1)(g-1-\n)} \;,\ee
where
\be z = e^{-2 \pi b \gamma}, \;\;\; \bar{z} = e^{-2 \pi b^{-1} \gamma } \;. \ee
Now the limit $\gamma \rightarrow \infty$ corresponds to taking both $z$ and $\bar{z}$ to $0$, and in this limit we have:
\be \eta(z) \approx e^{-\frac{\pi b \gamma}{12}} , \;\;\;  \eta(\bar{z}) \approx e^{-\frac{\pi b^{-1} \gamma}{12}}, \;\;\;\; \gamma \rightarrow \infty \;. \ee
Putting this together, we find a leading divergence of:
\be  \exp \bigg( -\frac{\pi \gamma}{12} (N-1) \bigg( (b +b^{-1})(g-1) + (b-b^{-1}) \n \bigg) \bigg) \nn \;,\ee
which agrees with \eqref{4dscal} in this limit.  It would  be interesting to extend this analysis to more general values of the flavor fugacity.

\subsubsection*{Punctures}

Above we derived a dual description for the 4d $\cN=1$ theory associated to the compactification of the 6d $\cN=(2,0)$ theory on a Riemann surface with flux, but with no punctures.  Let us briefly comment on extending the above result to the case with punctures.

After reducing on the $S^1$ factor, this corresponds to a codimension $2$ defect in the 5d $\cN=2$ SYM theory.  Then, as described in the context of 4d $\cN=2$ compactifications in \cite{Benini:2010uu,Chacaltana:2012zy,Yonekura:2013mya}, this defect can be understood by coupling the 5d degrees of freedom to the 3d $\cN=4$ theory $T_\rho[G]$ of \cite{Gaiotto:2008ak}, which describes a boundary condition of the 4d $\cN=4$ theory.  Then after performing the reduction above, this 3d theory will be coupled to the $G$ gauge group descending from the 5d gauge group.  The resulting theory will be a 3d $\cN=2$ star-shaped quiver-type theory, generalizing the $\cN=4$ star-shaped quiver of \cite{Benini:2010uu}.  It would be interesting to derive this also from the $S^3_b \times \Sigma_g$ partition function, but at present it is not known how to introduce such codimension $2$ defects in this observable.  Instead, in Section \ref{sec:duality}, we will describe how to incorporate punctures by starting with the known $\cN=4$ star-shaped quiver of \cite{Benini:2010uu}, which is known to be dual to the reduction of $\cN=2$ class ${\cal S}$ theories, and performing certain operations on both sides of the duality to obtain more general $\cN=2$ dualities.

\subsection{Global properties and higher form symmetries}
\label{sec:global}

One shortcoming of the above analysis is that, since we work on $\R^3$ or $S^3_b$, we are not sensitive to issues related to the global form of the 3d gauge group, \eg, whether it is $SU(N)$ or $SU(N)/\Z_N$.  This can be understood by carefully tracing the higher form symmetries of the 6d theory upon compactification.  Here we focus on the $\cN=(2,0)$ theory for concreteness, but we expect similar issues to arise for general $\cN=(1,0)$ SCFTs.

First, we recall that for a QFT with a $q$-form symmetry, $\Gamma$, the partition function, $Z_\omega$, depends on a choice of a cocycle $\omega \in H^{q+1}(\cM_d;\Gamma)$, which one can think of as a choice of background $(q+1)$-form gauge field coupled to the symmetry.  One may then gauge this symmetry by summing over values of this background field, introducing a ``dual'' background $(d-q-1)$-form gauge field, $\tilde{\omega}$, for the Pontryagin dual group, $\hat{\Gamma}$:
\be \tilde{Z}_{\tilde{\omega}} = \frac{1}{\sqrt{|\Gamma|}} \sum_{\omega \in \Gamma} e^{i \langle \omega,\tilde{\omega}\rangle} Z_{\omega} \ee
where we define the natural pairing $\langle \omega,\tilde{\omega}\rangle = \int_{\cM_d} \omega \wedge \tilde{\omega}$.  Performing  this operation again returns us to the original theory.

The 6d $\cN=(2,0)$ has a $2$-form symmetry, however, it has some subtle properties which are related to the fact that this is a {\it relative} QFT \cite{Freed:2012bs,Witten:2009at}.\footnote{See also \cite{Gang:2018wek,EKSW} for related issues in the context of the 3d-3d correspondence.}  Let us take the theory of some chosen ADE type, let $\Gamma$ be the corresponding abelian group (\ie, the center of the corresponding simply connected Lie group).  Then, as discussed in \cite{Witten:2009at,Tachikawa:2013hya}, we cannot consider an arbitrary background in $H^3(\cM_6,\Gamma)$, as above, but must first decompose:
\be H^3(\cM_6,\Gamma) \cong L \oplus \tilde{L} \ee
where $L,\tilde{L}$ are Lagrangian subgroups, \ie, the natural pairing on $H^3$ vanishes on each subgroup.  Now the partition function is replaced by a ``partition vector.''   For one choice of basis, the components can be labeled by element in $L$:
\be Z_\lambda, \;\;\;\; \lambda \in L \subset H^3(\cM_6,\Gamma) \ee
We may alternatively define:
\be \tilde{Z}_{\tilde{\lambda}} = \frac{1}{\sqrt{|L|}} \sum_{\lambda \in L} e^{i \langle \lambda,\tilde{\lambda}\rangle} Z_{\lambda} \ee
We see the two choices are essentially related by the duality mentioned above, so we might call this a ``self-dual $2$-form symmetry.''  Note also that the partition function is not well-defined by itself, only after making some choice of $L$ and $\lambda \in L$.

Next suppose we compactify a $d$-dimensional theory with a $q$-form symmetry $\Gamma$, on a manifold $\cC_p$.  In general we expect the compactified theory to have a $q$-form symmetry valued in $H_0(C_p;\Gamma) \cong \Gamma$, a $(q-1)$-form symmetry valued in $H_1(C_p;\Gamma)$, and so on up to a $(q-p)$-form symmetry valued in $H_p(C_p;\Gamma) \cong \Gamma$.  The operators which couple to these new symmetries come from the $q$-dimensional operators in the original theory partially wrapping the compactified directions, however, some of these may become very massive in the compactification limit, and the corresponding symmetries will then act trivially. 

Let us now see how  this works when we compactify the 6d theory to 3d.  Following the philosophy of this paper, we will consider the two possible compactification orders, going through 5d or 4d, and compare the results in each case.

\

\noindent {\it Compactifying to 5d first - } Here we expect to get the 5d $\cN=2$ SYM theory with simply connected ADE group $G$.  To be precise, the global form of the gauge group depends on our choice of $L$.  Note that $H^3(\cM_5 \times S^1)$ is isomorphic to $H^3(\cM_5) \oplus H^2(\cM_5)$, and each term is a Lagrangian subgroup.  Suppose we take $L$ to be the $H^3(\cM_5)$ subgroup.  Then the 5d partition function depends on a choice of $\omega \in H^3(\cM_5)$, which is the same data as an ordinary $2$-form symmetry \cite{Gang:2018wek}.  This is precisely the $G/\Gamma$ theory, which has such a $2$-form magnetic $\Gamma$ symmetry.  On the other hand, if we choose $L$ to be $H^2(\cM_5)$, we get the $G$ gauge theory, with a $1$-form electric $\Gamma$ symmetry.  Note that if the 6d $2$-form symmetry were an ordinary one, we would have both a $1$-form and $2$-form symmetry at the same time, but the self-duality implies we get only one at a time, and in fact the two choices are related by gauging.  We can think of these two choices as differing by whether we wrap the $3$-dimensional $2$-form charge operators of the 6d theory on the $S^1$, when they become the $2$-dimensional charge operators of the $2$-form magnetic symmetry in 5d, or leave them unwrapped, leading to the $1$-form electric symmetry charge operators.

Now we compactify further on a Riemann surface, $\Sigma_g$.  Then we have seen we get a star-shaped quiver, and the central node will just be the 5d gauge group.  Let us first consider the $G/\Gamma$ gauge theory.  Then the $2$-form symmetry decomposes into a $2,1$ and $0$-form symmetry in 3d, but we claim only the $0$-form symmetry survives in the low energy theory, becoming the magnetic $\Gamma$ symmetry of the 3d gauge theory. On the other hand, if we start with the $G$ gauge theory, we find only the $1$-form $\Gamma$ symmetry survives in 3d.  These can also be seen directly from 6d as the cases where we wrap the charge operators on the $S^1$, in the first case, or on the $\Sigma_g$, in the second.

\

\noindent {\it Compactifying to 4d first - } Alternatively, we can first compactify on $\Sigma_g$, obtaining a 4d class ${\cal S}$ theory.  Then we can write:
\be H^3(\cM_4 \times \Sigma_g) = & \nn \\ 
H^1(\cM_4)& \oplus (H^1(\Sigma_g)  \otimes H^2(\cM_4) )\oplus H^3(\cM_4) \nn \ee
Following \cite{Tachikawa:2013hya}, we may pick a Lagrangian subgroup, $K$, of $H_1(\Sigma_g)$, and this choice determines the $1$-form symmetry of the 4d theory, \eg, determining the global form of the gauge group.\footnote{For a genus $g$ Riemann surface there are $g$ pairs of $(a,b)$ cycles, and one can choose $K$ by choosing one of each of these, which corresponds to taking $G$ or $G/\Gamma$ for each $\cN=4$ loop (in an appropriate duality frame).} In addition, we should choose $L$ to include either $H^1(\cM_4)$, leading to a $0$-form $\Gamma$ symmetry, or $H^3(\cM_4)$, leading to a $2$-form $\Gamma$ symmetry.  

We claim the usual form of class ${\cal S}$ theories discussed in the literature corresponds to the former choice, and in particular, that these theories all come with a privileged $\Gamma$ $0$-form symmetry.  We will discuss the action  of  this symmetry more explicitly in the next section when we describe the 4d models in more detail.  If desired, we may gauge this symmetry to obtain a new theory with a $2$-form symmetry, corresponding to the other choice of subgroup.  These two choices correspond to taking the charge operators to be unwrapped, in the former case, or to wrap $\Sigma_g$, in the latter.

Now we compactify further to 3d.  If we started with the usual class ${\cal S}$ theory, with its $0$-form symmetry, we obtain a $0$-form symmetry in 3d.  We claim this is dual to the star-shaped quiver with $G/\Gamma$ gauge symmetry, as was already observed in \cite{Razamat:2014pta}.  We can see that in both cases, the 6d charge operators are wrapping only the $S^1$ we have reduced on.  On the other hand, if we start with the $2$-form symmetry in 4d, it will reduce to $2,1$ and $0$ form symmetries, but only  the $1$-form symmetry survives, and the theory is dual to the star-shaped quiver with $G$ gauge symmetry.  

On the other hand, the reduction of the $1$-form symmetry to 3d is more subtle.  As a simple example, if we started with a genus-one  Riemann surface, this may have either an electric or magnetic $1$-form symmetry depending on our choice of $K$ above, leading to either the $G$ $\cN=4$ SYM theory, or the electromagnetic-dual $G/\Gamma$ theory.  However, upon naive reduction to 3d, the former theory has only a $1$-form symmetry, and the latter only a $0$-form magnetic symmetry, even though the 4d descriptions are equivalent.  This apparent contradiction, which arises more generally whenever $g>0$, is presumably related to the fact that the naive reductions correspond to ``bad theories'' in the terminology of \cite{Gaiotto:2008ak}.  It would be interesting to understand this issue in more detail.

\section{Reduction of 4d class ${\cal S}$ theories}
\label{sec:4dreduction}

In this section we arrive at an alternative description of the 3d theories of the previous section.  Namely, starting from the 6d $\cN=(2,0)$ SCFT, we may compactify this on a Riemann surface $\Sigma_g$ with flux $\n$ for the $Sp(1)_F \subset Sp(2)_R$  symmetry to obtain a 4d $\cN=1$ theory.  We may then compactify this theory on a circle to obtain a 3d $\cN=2$ theory.  We expect the theory we obtain in this way to be equivalent to that obtained by first compactifying to 5d and then 3d, as in the previous section, and in this way we may derive  3d dualities, which  we consider in the following section.

\subsection{4d models}

\label{sec:4dmodels}

The theories associated to compactifications of the 6d $\cN=(2,0)$ theory on a punctured Riemann surface were originally discussed in \cite{Gaiotto:2009we} in the $\cN=2$ case, and later generalizations to $\cN=1$ theories were described in \cite{Benini:2009mz,Bah:2011je,Bah:2012dg,Agarwal:2015vla}.  We focus on the theories of type $A_{N-1}$.  To briefly summarize, the ingredients specifying the compactifications are as follows:

\begin{itemize}
	\item The genus, $g$, of the Riemann surface
	\item The number of punctures, $s$
	\item The Chern-degrees, $p$ and $q$, of the two line bundles describing the normal bundle of the M5 branes, constrained by:
	\be p+q = 2(g-1) +s \ee
	In the language of the previous section, where we took $s=0$, this corresponds to a $Sp(1)_F$ flux $\n$ of:
	\be \label{npq} \n = \frac{p-q}{2} \ee
	\item For each of the $s$ punctures, a choice of $SU(2)$ embedding, $\rho$, into $SU(N)$.
	\item Additionally, for each puncture, we choose a sign, separating them into $+$ punctures and $-$ punctures, and let $s_\pm$ denote the number of each type.
\end{itemize}

\begin{figure}
	\begin{center}
		\includegraphics[scale=0.48]{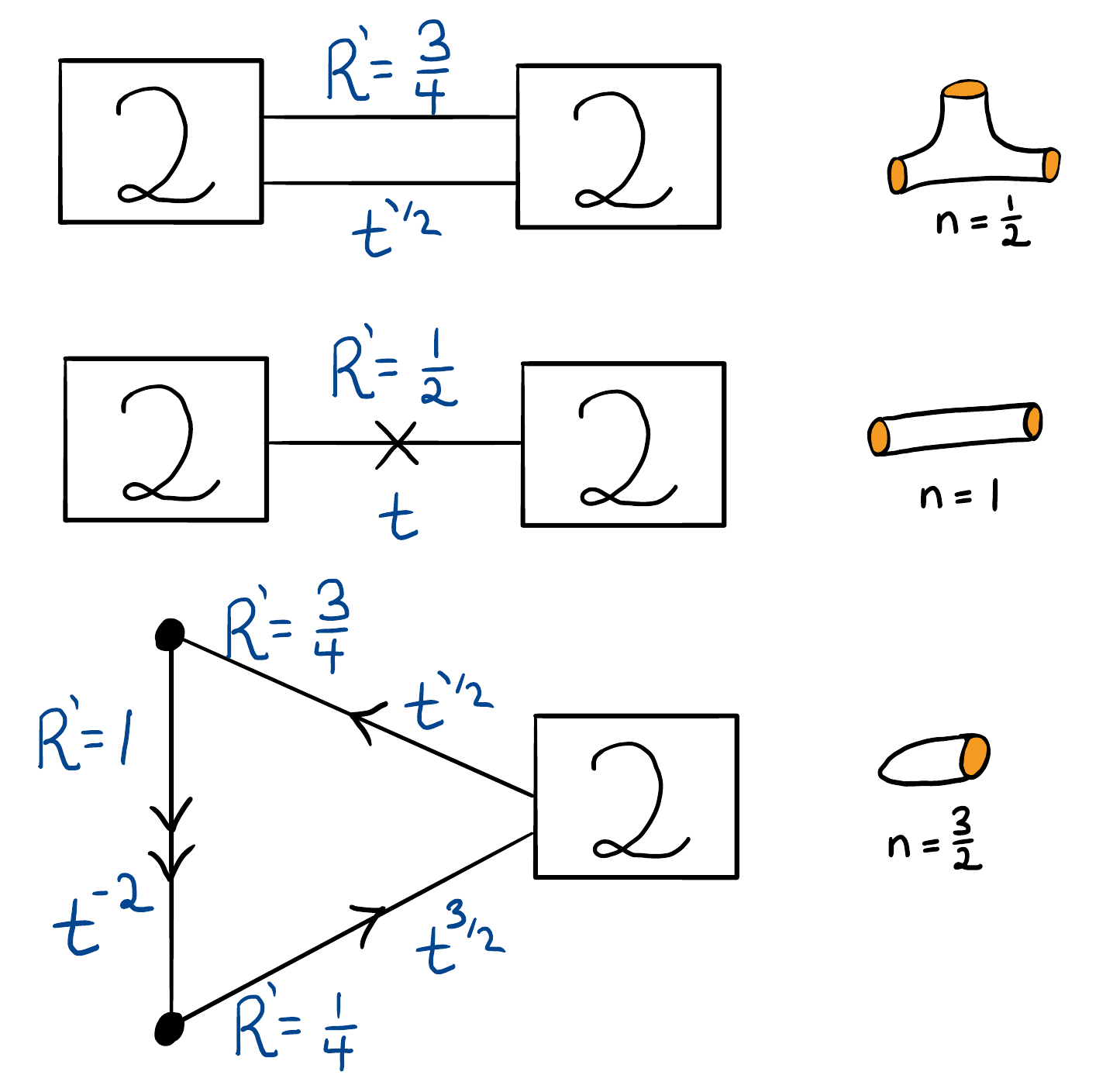}
		\caption{\label{fig:3bbA1}The building blocks from four dimensions. The powers of $t$ denote charge under $U(1)_t$, and $R'$ is the R-symmetry assignment. The double arrow for the theory on the bottom denotes having two fields of same charges. The $x$ denotes a chiral field flipping the $SU(2)^2$ quadratic invariant built from the bi-fundamental field.
		}
	\end{center}
\end{figure}

These theories are typically non-Lagrangian in the sense that there is no known simple Lagrangian describing the fixed point SCFT.  However, in the $A_1$ case, we may construct explicit Lagrangians for these models from three types of building blocks, shown in Figure \ref{fig:3bbA1}. The basic building block is a chiral field $Q_{ijk}$ in trifundamental irreducible representation of   $SU(2)\times SU(2)\times SU(2)$.
This theory corresponds to a sphere with three positive punctures. We assign to this chiral field charge $\frac12$ under the abelian global symmetry $U(1)_t$. We will also assign to it an R-charge of  $R'=\frac34$ for later convenience. This is not a superconformal R-symmetry but is useful for various computations. We also associate to the theory flux ${\frak n}=\frac12$.
Two other building blocks are obtained by closing the punctures \cite{Agarwal:2015vla}. We can obtain a theory corresponding to two punctured sphere with flux one by flipping and then closing one of the punctures.  We can construct a moment map operator ${M_i}^j=Q_{ilk}Q^{jlk}$, where indices are contracted with the epsilon symbol, and we introduce a field $\Phi$ in the adjoint representation of $SU(2)$ coupled to the moment map linearly, $Tr M \Phi$. Note that $\Phi$ has $R'$ charge $+\frac12$ and $U(1)_t$ charge $-1$. Next we give nilpotent expectation value to $\Phi$. This breaks the $SU(2)$ symmetry. In the IR the theory is a Wess-Zumino model built from a bifundamental chiral field of two $SU(2)$s and a flipper field. The bifundamental field has charge $-1$ under the $U(1)_t$ symmetry and R-charge $+\frac12$, with the flipper fields having R-charge $+\frac32$ and $U(1)_t$ charge $2$. We can further close an additional puncture to obtain the theory corresponding to a sphere with one puncture with flux $\frac32$. Closing punctures shifts the flux by $+\frac12$.

The procedure of flipping a puncture changes its sign. We can close a puncture without flipping by giving expectation value to the moment map. This will shift the flux by $-\frac12$. A general theory is obtained by gluing the blocks together. Gluing can be done \cite{Bah:2012dg} either by gauging diagonal $SU(2)$ symmetry of two punctures of same sign with introduction of adjoint field $\Phi$ coupled to the two moment maps, or by gluing two punctures of opposite sign turning on a superpotential coupling the two moment maps. These two procedures are consistent with the above definitions of signs of punctures.   

We will discuss explicit examples of these Lagrangians below, when we consider their dimensional reduction to 3d.

\subsection{General aspects of reduction to 3d}

We wish to reduce the four dimensional models on a circle to three dimensions. There are several general comments we want to discuss. First, as all the symmetries in four dimensions preserved by superpotentials are not anomalous in the theories we will consider, we do not expect to generate any monopole superpotentials upon reduction \cite{Aharony:2013dha}.  Second, we want to discuss the relation between marginal and relevant operators in four and three dimensions. If we have no $U(1)$ symmetries in four dimensions and no accidental symmetries appearing upon reduction, we expect that exactly marginal operators in four dimensions become exactly marginal in three and the same for relevant operators. However, in our setup we do have $U(1)_t$ symmetry. Upon reduction the superconformal R-symmetry will be in general different. Thus, the relevance of different operators might change. In particular a relevant operator can become marginal and a marginal operator can become either irrelevant or relevant. 

Let us discuss two examples. First consider the model corresponding to a tube with flux one. This is a Wess-Zumino model with cubic interactions in four dimensions and thus flows to a free theory. On the other hand the superpotential is relevant in three dimensions and the model flows to interacting SCFT. Thus we have an order of limits issue. If we first flow to the SCFT in four dimensions and then reduce to three we get a free model. If we first reduce on the circle with non vanishing superpotential and then flow we get an interacting theory. 
Another example is of giving mass to the adjoint chiral fields $\Phi$. If we build general models with three punctured spheres with flux $+\frac12$ we get theories with flux $g-1+\frac{s}2$, with $g$ being the genus and $s$ the number of punctures (which we take to be even). Upon giving masses to $\Phi$ we flow to a theory corresponding to same number of punctures and genus but with flux $0$. These are different models in four dimensions. The latter has quartic interactions for the fields and has a large conformal manifold not passing through zero coupling, with the former having cubic interactions and a conformal manifold passing through zero coupling. However, if we reduce the former model to three dimensions the superconformal R-symmetry assigns charge $+1$ to $\Phi$ and $+\frac12$ to $Q$ (see {\it e.g.} \cite{Bachas:2019jaa}). This means that a quadratic superpotential for $\Phi$ is marginal, as well as the quartic $Q^4$ and the cubic $\Phi Q^2$. The first two have opposite charges under $U(1)_t$ and thus a combination of them is an exactly marginal deformation \cite{Green:2010da}. Thus if we first reduce the model to three dimensions and then deform by a mass term for the adjoint we stay on the same conformal manifold. In particular although the two theories above are different in four dimensions they sit on same conformal manifold in three dimensions. In particular the models corresponding to different value of flux in four dimensions, $0$ and $g-1+\frac{s}2$, correspond to same model in three dimensions. 

\begin{figure}
	\begin{center}
		\includegraphics[scale=0.48]{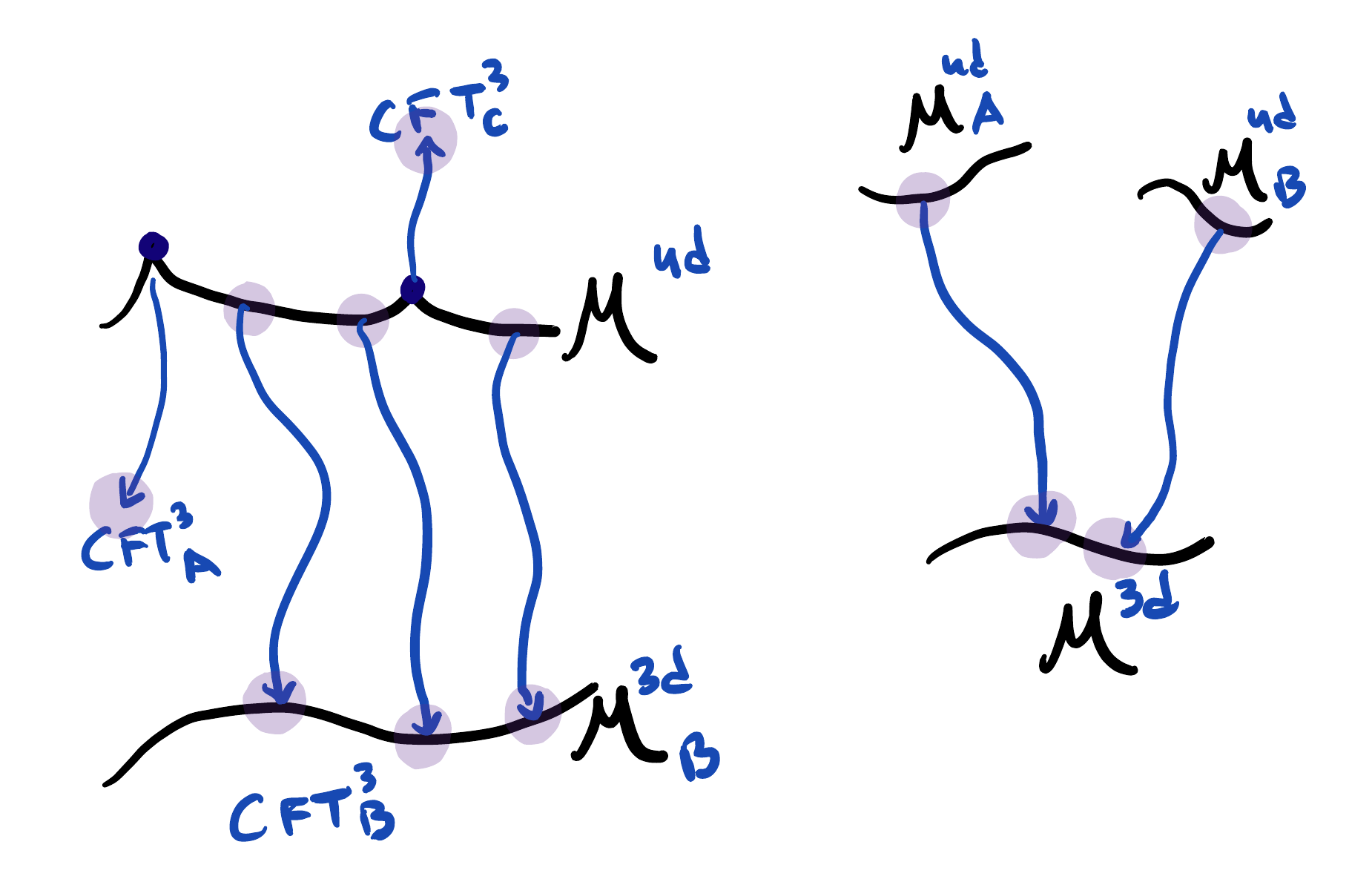}
		\caption{Different points on same conformal manifold might flow to different theories in three dimensions. Different models in four dimensions might flow to same conformal manifold in three dimensions.}
	\end{center}
\end{figure}

On the other hand, because a marginal operator in four dimensions can be relevant in three dimensions, two theories residing on the same conformal manifold in four dimensions can flow to different models in three. If we have a conformal manifold in four dimensions, for example which passes through zero coupling, the zero coupling cusp will flow to some CFT in three dimensions, but then turning on an operator which is marginal in four will be in general relevant in three dimensions and we will flow to a different model. In general we expect that a conformal manifold in four dimensions will reduce to different CFTs. The bulk of the manifold flowing to same CFT and various cusps or loci with enhanced symmetries flowing, possibly, to different models. 

These issues are important to us as we want to associate three dimensional models to Riemann surfaces with flux. As we have understood now we need to specify what we exactly mean by flux in three dimensions.  As far as compactification from six dimension goes the setup is that we take the $(2,0)$ theory on a Riemann surface times a circle and flow to an effective theory in three dimensions for which we then find a three dimensional UV completion. The  tunable free parameters in this setup are the relative sizes of the Riemann surface and the circle and the holonomy for the $U(1)_t$ symmetry. If we consider punctures then we also have holonomies for puncture symmetries around the circle. The ambiguity we encounter is related to the choice of these parameters. For special choices, say first tuning the surface to be zero size and then the circle to zero, we might get one answer while if we keep the parameters generic and finite, flow to effective three dimensional theory and then find UV completion, the answer can be different. In what follows we will suggest three dimensional models for the latter set-up.

\section{Duality}
\label{sec:duality}

In the previous two sections we have discussed the same system, the 6d $A_{N-1}$ $\cN=(2,0)$ SCFT compactified on $\Sigma_g \times S^1$ with flux, from two different points of view, leading to two different 3d descriptions.  In this section we discuss the relation between these description, and provide some checks that they are indeed equivalent. We will denote the reduction first to 5d and then to 3d as duality frame $B$, and the reduction first to 4d and then to 3d as duality frame $A$. To be concrete we will concentrate on the case of $A_1$ $(2,0)$ theory reductions as in this case all the models have simple Lagrangians.

\subsection{Genus zero compactifications with no punctures}

\begin{figure}
	\begin{center}
		\includegraphics[scale=0.52]{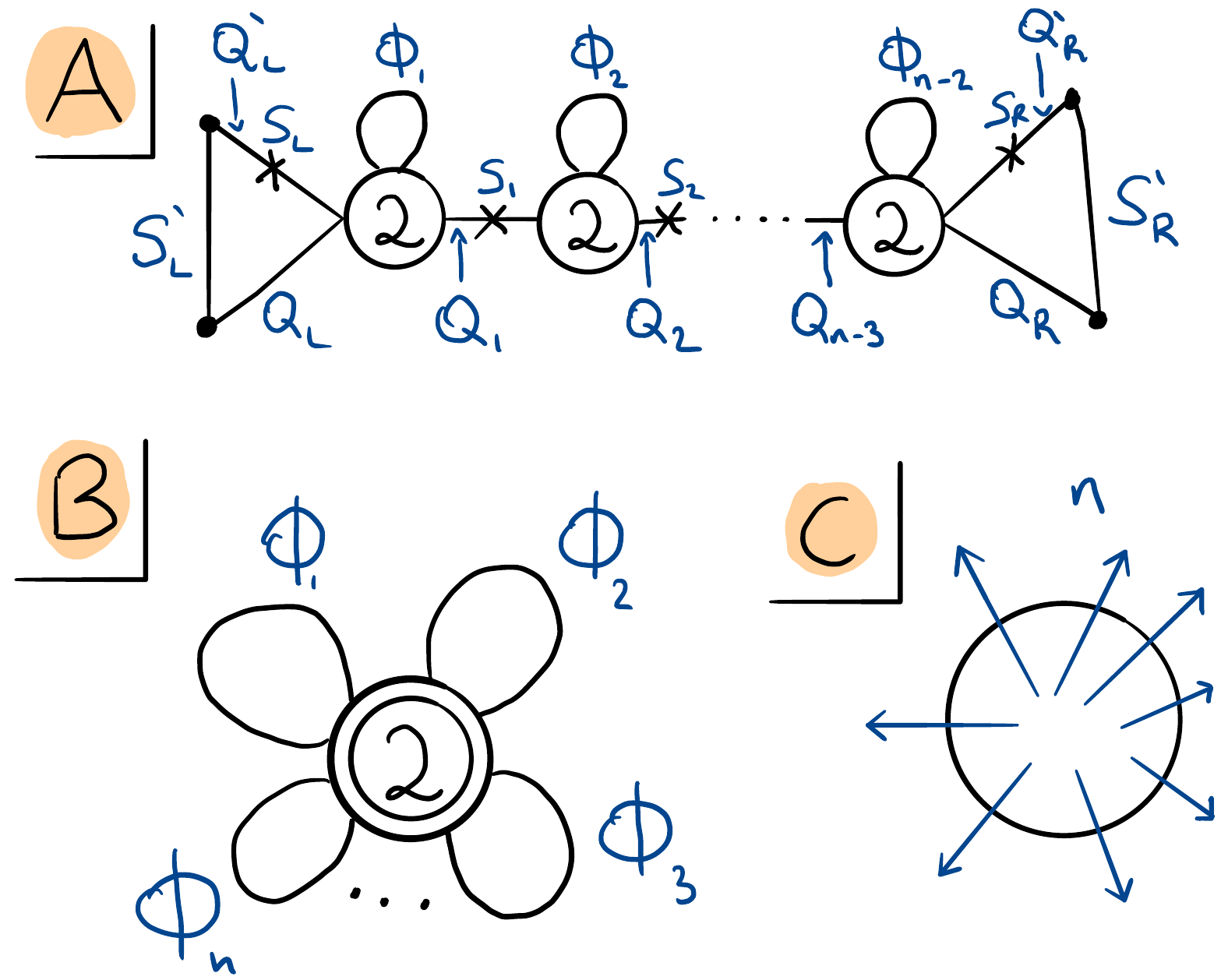}\label{figduals}
		\caption{Three descriptions of sphere with flux $\n$. $A$ is the quiver obtained by reducing  on a circle the 4d theory of class ${\cal S}$ corresponding to compactification on a sphere with flux $\n$. Here the fields $s_i$ are flip fields as in Figure \ref{fig:3bbA1}, while $s_L$ and $s_R$ are chiral fields coupling with superpotential as detailed in \eqref{LQsuperpot}. $B$ is the mirror dual, the theory obtained by reducing the 5d maximally supersymmetric Yang-Mills on a sphere with flux. Finally, $C$ is the reduction geometry. Here double circle denotes gauge group $SU(2)/{\mathbb Z}_2$ while a circle the group $SU(2)$.    }
	\end{center}
\end{figure}

Let us first consider the example of a sphere with no punctures and flux $\n$.  The theory on side $B$ is given by $SU(2)/{\mathbb Z}_2$ gauge theory with $\n$ adjoints.\footnote{The fact that the gauge group is $SU(2)/{\mathbb Z}_2$ and not $SU(2)$ is important for the duality \cite{Razamat:2014pta}. The various partition functions one can compute depend on the global structure of the gauge groups and the difference between  $SU(2)/{\mathbb Z}_2$  and $SU(2)$ can be easily detected leading to some discrepancies if the wrong choice is made, as was first noticed in \cite{Benvenuti:2011ga}.}  There is no superpotential.   On side $A$ of the duality we build the models combining together the blocks of Figure \ref{fig:3bbA1}. For flux ${\frak n}>2$ (we will comment on ${\frak n}=2$ soon) we obtain a model with $\n-2$ $SU(2)$ gauge groups depicted in Figure \ref{figduals}. The superpotential is,
\bea \label{LQsuperpot}
& \ds W = \sum_{i=1}^{\n-3} (Q_i^2\Phi_i\Phi_{i+1}+Q_i^2 s_i)+Q_L \Phi_1 Q_L+Q'_L\Phi_1 Q'_L s_L \nn \\
& \ds+Q_L Q'_L s'_L + Q_R\Phi_{\n-2} Q_R+ Q'_R\Phi_{\n-2} Q'_R s_R  +Q_R Q'_R s'_R\,.\nn\\
\eea Note that the fields $s_L, s_L',s_R$, and $s_R'$ have the same charges. This superpotential is the most general one consistent with the charges of various fields. The charges of the various fields appear in Figure  \ref{fig:3bbA1} with $Q_{L,R}$ having $R'$ charge $\frac34$, $Q'_{L,R}$ charge $\frac14$, and $s_{L,R}$, $s'_{L,R}$ $R'$ charge $1$. The $R'$ charge of the adjoint fields on side $B$ is $\frac12$, the $t$ charge is $-1$, and there is no superpotential.
Several comments are in order. The theory on side $B$ has a manifest $U(\n)$ symmetry which is not apparent on side $A$. This symmetry is conjectured to emerge in the IR. Second, the theory on side $B$ has a manifest ${\mathbb Z}_2$ symmetry under which the monopoles which have proper quantization for $SU(2)/{\mathbb Z}_2$ but not $SU(2)$ are charged. The symmetry is manifest also on side $A$. Note that we build theories on side $A$ from $\n-1$ building blocks of  Figure \ref{fig:3bbA1}, each containing fields charged under the gauged symmetries and gauge singlet fields. We can consider the ${\mathbb Z}_2$ symmetry under which the fields transforming under gauged symmetries in only one block are charged. This is exactly the ${\mathbb Z}_2$ symmetry we need to identify. Note that it does not matter which block is chosen as different choices are related by gauge transformations of the different $SU(2)$ gauge symmetries. Using the ${\mathbb Z}_2$ center of the gauge groups we can redefine the block which is charged under the symmetry and in more generality any odd number of blocks can transform under the ${\mathbb Z}_2$ symmetry. 

Let us discuss the map of operators. The basic operators on side $B$ of the duality are traces of quadratic combinations of the adjoints. This is in the two index symmetric representations of $U(\n)$.  On side $A$ the operators which build this are the $\n+1$ flip fields, $\n-2$ adjoint quadratic,   and there are $\frac12(\n-2)(\n-1)$ monopole operators. The GNO charges are as follows. We have $\n-2$ gauge groups with the first and last having different content than others. The charges are, $\n-2$ basic monopoles $(0,\cdots,1,\cdots)$, the monopole, $(1,1,1,\cdots, 1 )$, monopoles of form $(0\cdots ,1,\cdots,1,\cdots)$ excluding $(1,0\cdots, 0,\cdots,1)$. There is no obvious symmetry acting on these operators but we claim they form the symmetric two index representation of $U({\frak n})$ symmetry conjecturally emerging in the IR.

We can perform $Z-$extremization to determine the superconformal R-symmetry. The R-symmetry increases with flux starting from slightly above $1/4$ for $\n=2$ and approaching $1/2$ as flux goes to infinity. For $\n>2$  there are no operators below unitarity (for $\n=2$  see below). 
In particular the cubic and quartic superpotentials that one can turn on are relevant deformations. All operators are above the unitarity bound.  Let us now discuss a couple of examples in more detail.

\subsubsection*{The case of $\n=2$}

\begin{figure}
	\begin{center}
		\label{fign2}\includegraphics[scale=0.54]{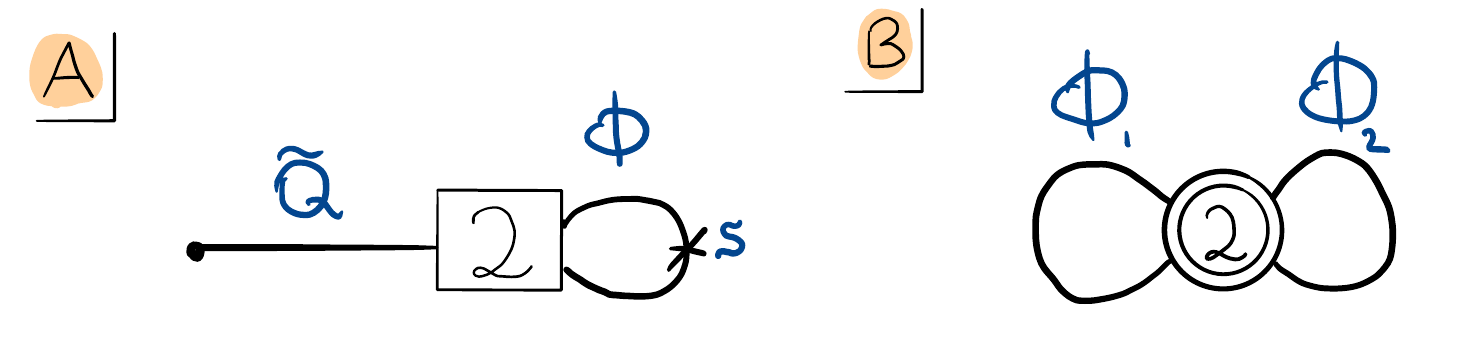}
		\caption{Sphere with ${\frak n}=2$. The theory on side $A$ is a WZ model with  superpotential appearing in \eqref{f2sup}.}
	\end{center}
\end{figure}

The case of $\n=2$ is a bit special as it is not built by gluing together building blocks of Figure \ref{fig:3bbA1} but rather by taking the sphere with one puncture and flux $\frac32$ of that figure, flipping the puncture and closing it. The resulting theory is a WZ model in Figure \ref{fign2}.
 On  side $A$ we thus have the WZ model with six fields and superpotential, 
\be
\label{f2sup}W =  \widetilde Q \Phi \widetilde Q+{\frak s}^2\, Tr\Phi^2\,.
\ee As this is a WZ model with quartic superpotential we need to be careful whether it is interacting and indeed ${\frak s}$ goes below unitarity bound. On side $B$ it is dual to a monopole operator. 
On  side $B$ we have SQCD with gauge group $SU(2)/{\mathbb Z}_2=SO(3)$ and two adjoint fields. The symmetry is $U(2)$. 
This is the only case where we can see explicitly the symmetry on side $A$. The six fields are organized into adjoint, fundamental, and a singlet. There is a cubic superpotential between the fundamental squared and the adjoint, as well as quartic between adjoint squared and singlet squared. This is the same duality as the well known one between $SO(N)$ and $SO(N_f-N+2)$ with $N=3$ and $N_f=2$, see \cite{Aharony:2013kma} which is the $SO$ version of Aharony duality \cite{Aharony:1997gp}.  

\subsubsection*{The case of $\n=3$}

Let us consider the special case of $\n=3$.  The theory has the same matter content as $SU(2)/{\mathbb Z}_2$ ${\cal N}=8$ SYM however we do not turn on the cubic superpotential. The dual model is just an $SU(2)$ gauge theory with two flavors and an adjoint coupled with a quartic superpotential and four gauge singlet fields.  
We can consider the supersymmetric index of the two dual frames \cite{Kim:2009wb,Kinney:2005ej,Willett:2016adv}. The index is given by,
\be\label{defindex}
{\cal I}={\text Tr}_{S^2} (-1)^{2J_3} q^{\frac12(E+J_3)} \prod_{i=1}^{rank \, G_F} a_i^{q_i}\,. 
\ee Here $J_3$ is the Cartan of the $SO(3)$ Lorentz symmetry, $E$ is the scaling dimension, $R$ is the R-charge, $G_F$ is the global symmetry group, and $q_i$ are the charges under the Cartan of $G_F$. The trace is over states in radial quantization. States contributing to the index satisfy $E-J_3-R=0$.

When computed in expansion in $q$ at order $q^1$ the index captures the marginal operators minus the conserved currents \cite{Razamat:2016gzx,Beem:2012yn}. In  particular it then captures 
the number of exactly marginal deformations by applying the procedure of  \cite{Green:2010da}.
The index of the two sides of the duality at order $q$ is given by,
\be
&&A:\; 1\;\\
&&B:\; {\bf 10}-{\bf 8}-1\,.\nonumber
\ee  On side $B$ we refine the index with $SU(3)$ fugacities, which we cannot do on side $A$ as this symmetry is only emergent in the IR. Here ${\bf 8}$ stand for the character of the adjoint representation of $SU(3)$ and ${\bf 10}$ for the character of the three index symmetric representation.
Note that this implies that the theory has a conformal manifold which has two complex dimensions. In description $A$ this is given by the index at order $q$ plus one, as at this order the index counts marginal minus conserved current operators and the $U(1)_t$ symmetry is not broken on the conformal manifold. In description $B$ we have marginal operators in ${\bf 10}$ of the $SU(3)$ symmetry group. We  can perform the computation of counting the dimension of the conformal manifold by counting the number of holomorphic invariants \cite{Green:2010da} built from marginal couplings which for ${\bf 10}$ is two. On conformal manifold thus the $SU(3)$ symmetry is broken and the dimension can be understood as $10-8=2$.\footnote{ Let us here mention a cute observation. We can study the theory here deformed by flipping the sextet of operators $Tr \Phi_{\{i}\Phi_{j\}}$. This is a relevant deformation. Interestingly doing so we find that at order $q$ we now have $-2 \times {\bf 8}-1$. This indicates that the symmetry is enhanced to $SU(3)\times SU(3)\times U(1)_t$. In particular the second $SU(3)$ comes from the monopoles of $SO(3)$ which are not properly quantized for $SU(2)$. This means that the $SU(2)$ theory does not have this enhancement. The symmetry we see in the UV is the diagonal combination of the symmetries. This enhancement of symmetry is very reminiscent of the ones discussed in \cite{Razamat:2018gbu,Sela:2019nqa} in 4d where it was understood in terms of reductions of 6d models. It would be interesting to understand whether a similar type of explanation can be found here.}

\

\

In general, we can start on side $B$ of the duality with flux $\n$ and obtain any lower flux by giving masses to some of the adjoint fields. On side $A$ this can not be done as giving mass to one of the adjoint will generate masses to all and also linear superpotentials to the flip fields and some of the monopoles. The $SU(\n)$ emerges in the IR and to be able to give mass only to some of the operators related by the symmetry we need to be at the fixed point.

\subsection{Adding punctures and handles}

\begin{figure}
	\begin{center}
		\includegraphics[scale=0.54]{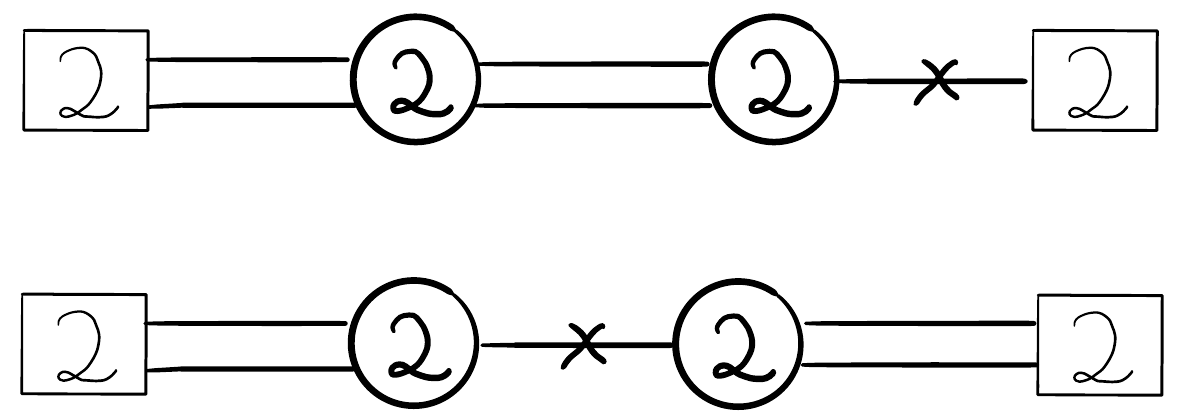}  \label{fig:duexa}
		\caption{An example of a duality. Combining different building blocks or combining same blocks in different order should give same CFT.}
	\end{center}
\end{figure}

Next we discuss surfaces with punctures and handles. Let us first discuss side $A$ of the duality. 
  When we add punctures  we introduce a new ingredient into the field theoretic construction. On side $A$ we consider quivers as above but we also allow some of the links between gauge groups to be tri-fundamental fields, see Figure \ref{fig:3bbA1}. Each such field will have $SU(2)$ flavor symmetry corresponding to a puncture.  
We have several building blocks (see Figure \ref{fig:3bbA1}) and we can construct same models using different blocks and combining them in different order. Each block is associated to some value of flux and as long as number of punctures and the flux are the same two theories should correspond to same IR CFT, see for example Figure \ref{fig:duexa}.  Reducing these models to three dimensions we obtain ``good'' theories in the nomenclature of  \cite{Gaiotto:2008ak}. Then to construct theories corresponding to compactifications with handles in four dimensions we just glue pairs of punctures together. Gluing corresponds as before to gauging a diagonal combination of puncture $SU(2)$ symmetries and introduction of adjoint field $\Phi$ coupled through superpotentials to moment maps.
However, these theories upon reduction to three dimensions are ``bad''. In particular their partition functions do not converge. This is believed to be due to the fact that the superconformal R-symmetry cannot be identified correctly from the UV description. This is not to claim that theories corresponding to compactifications  on surfaces with handles have some intrinsic problem, just that the direct reduction of the 4d theories is not a useful way to discuss them. A well known example of this is the ${\cal N}=8$ SYM in 3d, a useful description of which is the ABJM CS/matter theory \cite{Aharony:2008ug}.  The description $B$ we will construct will be good also in presence of handles.
In particular, we will be thus able to check dualities between side $A$ and $B$ only for genus zero compactifications, albeit with any number of punctures and any value of the flux.

To consider theories on side $B$ we need to discuss a new building block, which is the well studied $T[SU(2)]$ theory, which is a special case of the $T[SU(N)]$ theory of Gaiotto and Witten \cite{Gaiotto:2008ak}.  Let us start by reviewing some relevant properties of this theory, focusing in particular on the special case $N=2$. More useful observations about these models are discussed in Appendix \ref{app:tsun}.

\subsubsection*{Properties of $T[SU(N)]$}

The $T[SU(N)]$ theory is a 3d $\cN=4$ model which describes a domain wall interpolating between $S$-dual instances of the 4d $\cN=4$ SYM theory \cite{Gaiotto:2008ak}.  It has $SU(N)_H \times SU(N)_C$ flavor symmetry, where we have identified the factors as the ``Higgs'' and ``Coulomb'' symmetries, in addition to the $U(1)_t$ symmetry of the $\cN=2^*$ algebra.

There are several known descriptions of this model manifesting different subsets of the global symmetry, see Appendix \ref{app:tsun}. In the $N=2$ case, this model has a description as a $U(1)$ ${\cal N}=4$ gauge theory with two charge one hypermultiplets. The model has manifest ${\cal N}=4$ supersymmetry and $SU(2)_H\times U(1)_C$ flavor symmetry.\footnote{The $SU(2)_H \times SU(2)_C$ flavor symmetry in the $N=2$ case should not be confused with the $\cN=4$ R-symmetry, which we do not discuss in this subsection.}  The $U(1)_C$ symmetry is the topological symmetry coming from the gauge abelian symmetry and $SU(2)_H$ rotates the two hypermultiplets. It is conjectured that the $U(1)_C$ symmetry enhances to $SU(2)_C$. Thus the symmetry in the IR is $SU(2)_H\times SU(2)_C$ and the model is self dual under mirror symmetry. That is one $SU(2)$ rotates the Coulomb branch and the other the Higgs branch with the two branches isomorphic.  More generally, $T[SU(N)]$ has a description as a triangular quiver gauge theory, with gauge group $U(1) \times U(2) \times ... \times U(N-1)$, with bifundamental hypermultiplets between adjacent gauge groups, and $N$ flavors in the final $U(N-1)$ gauge groups, leading to a manifest $U(1)_C^{N-1} \times SU(N)_H$ flavor symmetry, which is enhanced in the IR to $SU(N)_H \times SU(N)_C$.  

To discuss the structure of the flavor symmetry, it is convenient to schematically write the partition function on an unspecified manifold, coupling the flavor symmetries to background vector multiplets $\cV_H,\cV_C$ and $\cV_t$, as:
\be Z_{T[SU(N)]}(\cV_H,\cV_C,\cV_t) \;. \ee
Then this partition function is self dual under mirror symmetry, namely:
\be \label{tsunid1} Z_{T[SU(N)]}(\cV_H,\cV_C,\cV_t) = Z_{T[SU(N)]}(\cV_C,\cV_H,-\cV_t)\;. \;\;\;\;\; \ee
In addition, following \cite{Aprile:2018oau}, it is convenient to define a related theory called $FT[SU(N)]$, by ``flipping'' one of the $SU(N)$ flavor symmetries, say $SU(N)_C$.  This means we couple an adjoint chiral multiplet to the moment map  for this symmetry, which fixes this adjoint to have $U(1)_t$ charge $-1$, and we denote this as
\be \label{ftsun} & \ds Z_{FT[SU(N)]}(\cV_H,\cV_C,\cV_t) \hspace{3cm}  \\
& \ds=  Z_{T[SU(N)]}(\cV_H,\cV_C,\cV_t) Z_{adj}(\cV_C,-\cV_t) \;. \nn \ee
Then the $FT[SU(N)]$ theory is symmetric in its two flavor symmetries, without performing a $U(1)_t$ conjugation, \ie:
\be \label{ftsunid}Z_{FT[SU(N)]}(\cV_H,\cV_C,\cV_t) =  Z_{FT[SU(N)]}(\cV_C,\cV_H,\cV_t) \;. \nn \ee
Note this is an independent identity from \eqref{tsunid1}.  It can be rearranged into the following statement for the $T[SU(N)]$ theory:
\be \label{tsunid2} & \ds Z_{adj}(\cV_H,\cV_t) Z_{T[SU(N)]}(\cV_H,\cV_C,\cV_t) Z_{adj}(\cV_C,-\cV_t) \nn \\
 &= \ds Z_{T[SU(N)]}(\cV_H,\cV_C,-\cV_t)  \ee
This says that flipping both symmetries of $T[SU(N)]$ gives the same theory up to an overall $U(1)_t$ conjugation  \cite{Aprile:2018oau}.  Equivalently, by \eqref{tsunid1}, this is the same as exchanging the two $SU(N)$ background gauge fields.

\subsubsection*{$\cN=2$ star-shaped quivers}

Using the $T[SU(N)]$ block we can now state what is the dual on side $B$ of theory with flux $\n$ and $s$ punctures.  We claim the dual of the general 3d $\cN=2$ class ${\cal S}$ theory, with data as described in Section \ref{sec:4dmodels} above, is given as follows (specializing to the case of all maximal punctures, for simplicity).

\begin{itemize}
	\item A central gauge group $SU(N)/\Z_N$.
	\item This gauge group has:\footnote{We recall from \eqref{rtdef} that $U(1)_F$ is essentially the charge conjugation of the symmetry $U(1)_t$ discussed in the 4d models and their reduction above.  More precisely, for $g>0$ there is an additional admixing with a hidden symmetry, but we will  mostly consider the case $g=0$ below.} 
	\begin{itemize}
		\item $\ell$ adjoint chirals, $\Psi$, of $U(1)_F$ charge $1$ and R-charge $1$,
		\item $\tell$ adjoint chirals, $\widetilde\Psi$, of $U(1)_F$ charge $-1$ and R-charge $1$, and
		\item $g$ adjoint chirals, $\Omega$, which have zero charge under both $U(1)_F$ and $U(1)_R$.
	\end{itemize}
	As in the case without punctures, we cannot in general specify $\ell,\tell$ individually, however, we claim their difference satisfies:\footnote{That the RHS is an integer follows from the restriction on $p$ and $q$ above.  Explicitly, it can also be written as $g-1-q +s_-$ or $1-g+p-s_+$.}
	\be \ell - \tell = \frac{p-q-s_++s_-}{2} \label{lplm} \ee This is because the above charges are consistent with superpotential terms of the form $\Psi \widetilde \Psi$, which are mass terms.
	\item For each puncture, we couple a $T[SU(N)]$ theory to this central gauge group.  More precisely, for a positive puncture, we couple this to the Higgs symmetry of $T[SU(N)]$, leaving the Coulomb symmetry as a flavor symmetry, and for a negative puncture we couple to the Coulomb symmetry, leaving a Higgs flavor symmetry.  For this reason we will identify the Coulomb and Higgs symmetries of $T[SU(N)]$ as corresponding to ``positive'' and ``negative'' puncture flavor symmetries, respectively.
\end{itemize}

\begin{figure}
	\begin{center}
		\includegraphics[scale=0.68]{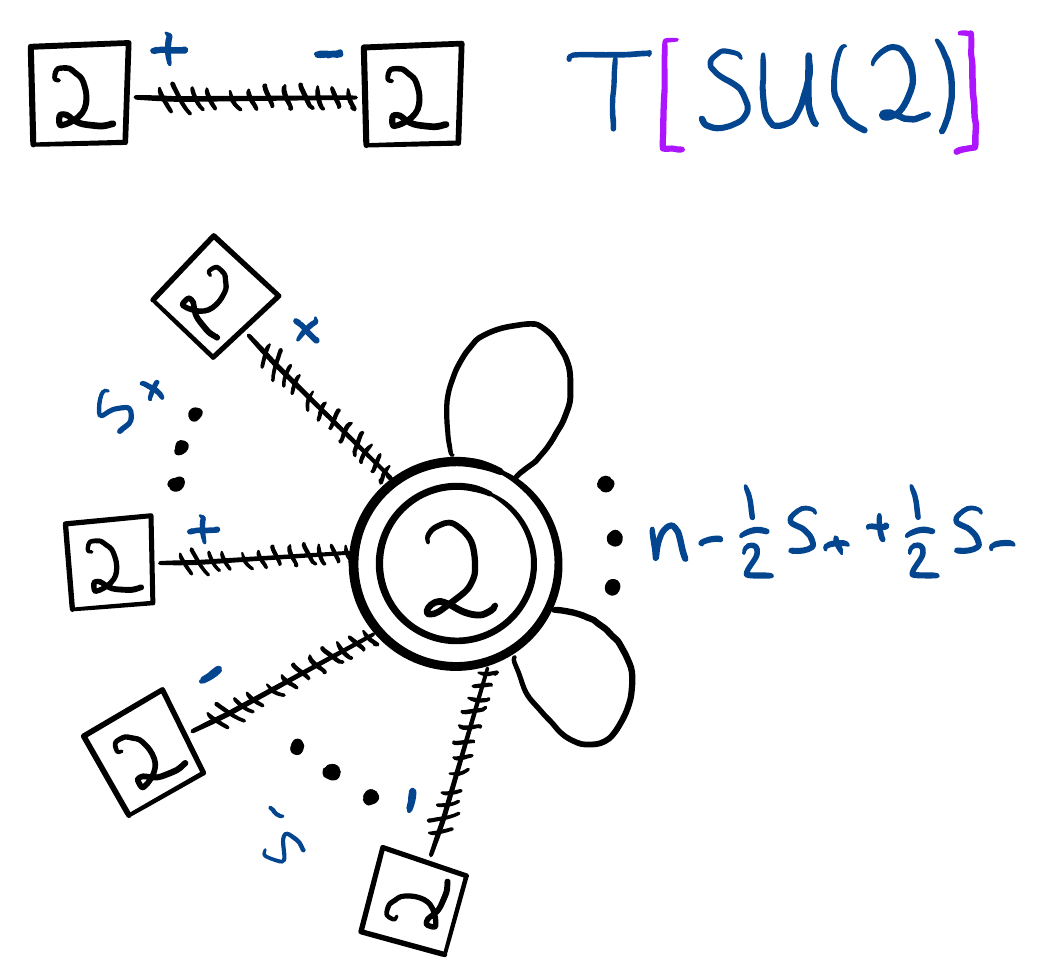}
		\caption{Above is an abstract depiction of $T[SU(2)]$. The $+$ and $-$ refer, respectively, to the Coulomb and Higgs flavor symmetries of the theories. On the bottom the sphere with flux $\n$. The number of positive punctures is $s_+$ and negative is $s_-$. }
		\label{fig:ssqquiv}
	\end{center}
\end{figure}

This is illustrated in Figure \ref{fig:ssqquiv} in the case $N=2$.  Let us now go through several consistency checks of this formula.

\begin{enumerate}
	\item {\bf Zero puncture case} - In the case with zero punctures and flux $\n$,  comparing to \eqref{3dssqs0}, and recalling $\n=\frac{p-q}{2}$, we see this reproduces our description derived above by reduction from 5d as an $SU(N)/\Z_N$ gauge theory with adjoints of various $U(1)_F$ charges.
\item {\bf $\mathbf{3d \; \cN=4}$ case} In the special case:
\be p = 2(g-1) + s, \;\;\; q=0, \;\;\;s_+=s, \;\;\;s_-=0 \ee
the theory has 3d $\cN=4$ supersymmetry.  In this case, there is a known SSQ ``mirror''\footnote{To avoid confusion, we work with a fixed definition of $U(1)_t$  on both sides of the duality, rather than mapping it with a change of sign as usual in such a mirror symmetry.  Then the $U(1)_t$ charges in the SSQ description will have opposite sign compared to the usual conventions in the literature.} dual description \cite{Benini:2010uu}.  This is given by an $SU(N)/\Z_N$ central node with an adjoint chiral multiplet of $U(1)_t$ charge $1$ (in the $\cN=4$ vector multiplet), and $g$ adjoint hypermultiplets.  The latter naturally have $U(1)_t$ charge $-1/2$, but as in \eqref{hidmix}, we admix this with the $U(1)$ flavor symmetry rotating all of the hypers so that half of the adjoint chirals have charge $-1$ and half have charge zero.  Then we find $g$ adjoint chirals with $U(1)_t$ charge zero, along with $\ell=g$ of charge $-1$ and $\tell=1$ of charge $1$.  Recalling that $T=-F$, we see this indeed agrees with \eqref{lplm} in this case, which gives:
\be \ell-\tell = g-1 \ee
These are all coupled to $s$ copies of the $T[SU(N)]$ theory, which are oriented so that the `$-$' (Higgs) symmetries are facing inward, as above.

There is another special case which gives an $\cN=4$ theory, which is:
\be p = 0, \;\;\; q= 2(g-1) + s, \;\;\;s_+=0, \;\;\;s_-=s \ee
Then we see the description is much as above, except that now:
\be \ell-\tell = 1-g \ee
so that we may take the adjoint chirals to have opposite $U(1)_t$ charge.  In addition, the $T[SU(N)]$ symmetries are now oriented so that their `$+$' (Coulomb) symmetries are facing inward, or equivalently, using \eqref{tsunid1}, the Higgs symmetry faces inwards but we apply a $U(1)_t$ conjugation.  Then, using \eqref{tsunid2}, this is precisely the same theory as above up to an overall $U(1)_t$ conjugation.  This is indeed the expected result, as such a conjugation exchanges the role of $p$ and $q$  and of `$+$' and `$-$' punctures.

\item {\bf Flipping punctures}

The operation of flipping a puncture of `$+$' type (respectively, `$-$' type) corresponds to introducing a new adjoint chiral multiplet of $U(1)_t$ charge $-1$ (respectively, $+1$), which couples to the moment map of that  symmetry.  Suppose we flip a `$+$' puncture in the above description.  From \eqref{tsunid2}, if we were to also flip the other puncture, this would be the same as reversing the orientation of the $T[SU(N)]$ theory.  Then this operation has the effect of reversing the $T[SU(N)]$ leg and adding a charge $T=-1$, or $F=1$, adjoint chiral to the central gauge node, which shifts:

\be \ell - \tell \;\;\; \rightarrow  \;\;\; \ell - \tell + 1 \ee
But this gives precisely the expected SSQ for this new theory, which now has $(s_+,s_-) \rightarrow (s_+-1,s_-+1)$.  A similar argument applies to `$-$' punctures.

\item {\bf Gluing punctures}

Next we consider the operation of gluing punctures.  For this we first need to mention another property of the $T[SU(N)]$ theory.  If we take two consecutive copies and glue them with an $\cN=2$ vector multiplet, we find a ``delta-function theory:''
\be \label{tsunglue} &\ds \int {\cal D} a_2 \; Z_{T[SU(N)]}(a_1,a_2,\tau) Z_{T[SU(N)]}(a_2,a_3,\tau) \nn  \\
& \ds= \delta(a_1-a_3) \ee
Here a ``delta function theory'' is a formal functional of the background fields which acts as a delta function in the path-integral, identifying the two gauge fields when one of them is path-integrated over.  This property of the $T[SU(N)]$ theory is expected on general grounds from its role as an $S$-duality wall, and has been verified in certain supersymmetric partition functions \cite{Benvenuti:2011ga,Nishioka:2011dq,Razamat:2014pta}.

Now suppose we have two SSQ's, as above, and we glue a `$+$' puncture of one to a `$-$' puncture of the other.  The general rules  for gluing class ${\cal S}$ theories \cite{Benini:2009mz,Bah:2011je} tell us we should use an $\cN=2$ vector multiplet, and then using \eqref{tsunglue}, we see that the two central nodes collapse to one, which contains all of the $T[SU(N)]$ legs and adjoint chiral multiplets of the two original quivers.  There are no additional adjoints introduced into the central node.  Thus we find:
\be \ell^{new} - \tell^{new} = \ell^{(1)} -  \tell^{(1)} +  \ell^{(2)} -  \tell^{(2)} \ee
This  is indeed the expected result from \eqref{lplm}, since the values of $p$ and $q$ of the two quivers simply add, and since we remove one `$+$' and one `$-$' puncture  in the gluing, the difference $s_+-s_-$ in \eqref{lplm} does not change.

If, on the other hand, we glue two `$+$' punctures, we should now gauge using a 3d $\cN=4$ vector multiplet.  Then, using \eqref{tsunid1} and \eqref{tsunid2}, we find:
\be  \label{tsunglueadj} & \ds \int {\cal D} a_1 \; Z_{T[SU(N)]}(a_1,a_2,\tau) Z_{T[SU(N)]}(a_1,a_3,\tau) Z_{adj}(a_1,-\tau) \nn \\
& \ds = \delta(a_2-a_3)Z_{adj}(a_2,-\tau)  \ee
This says that if we gauge the two `$+$' punctures together with an $\cN=4$ vector multiplet (\ie, including an additional adjoint chiral of $U(1)_t$ charge $-1$), then we obtain a delta function theory with an additional adjoint chiral multiplet of $U(1)_t$ charge $-1$.  Thus we now find:
\be \ell^{new} - \tell^{new} = \ell^{(1)} -  \tell^{(1)} +  \ell^{(2)} -  \tell^{(2)} + 1 \ee
Once again, this is the expected result, since now we have reduced $s_+$ by two.  A similar statement holds for gluing two `$-$' punctures.

\end{enumerate}

\section{Partition function checks}
\label{sec:pfchecks}

In this section we outline several computations of supersymmetric partition functions which lend further evidence to many of the dualities and relations we discussed above.  

\subsection{Supersymmetric index}

Let us detail some of the supersymmetric index checks of the dualities presented above. The index is given by the trace formula \eqref{defindex}. The technology of computing it for gauge theories has been developed in \cite{Kim:2009wb} and here we will follow the notations of \cite{Aharony:2013kma}. The index of a chiral superfield is given by \cite{Razamat:2014pta},
\be
&&{\cal I}_R (z,{\bf m};q)=\\
&&\qquad \left(q^{\frac{1-R}2}z^{-1}\right)^{\frac{\bf m}2} \prod_{i=0}^\infty\frac{1-(-1)^{\bf m}z^{-1}q^{\frac12|{\bf m}|+1-\frac{R}2+i}}{1-(-1)^{\bf m}zq^{\frac12|{\bf m}|+\frac{R}2+i}}\,.\nonumber
\ee Here $R$ is the R-charge, $z$ is a fugacity for $U(1)$ symmetry under which the superfield is charged and ${\bf m}$ is the magnetic flux for this symmetry through the $S^2$.
The index of the $T[SU(2)]$ model is then given by,
\be
&&{\cal I}(w,{\bf n}|z,{\bf m};q,t)={\cal I}_1(t^{-1},0;q)\sum_{l\in{\mathbb Z}+\epsilon(m)} w^{2l}\oint \frac{dh}{2\pi i h} h^{2{\bf n}}\nonumber\\
&& {\cal I}_{\frac14}(t^{\frac12} (h z)^{\pm1};\pm({\bf l}+{\bf m}) ;q)  {\cal I}_{\frac14}(t^{\frac12} (h^{-1} z)^{\pm1};\pm({\bf m}-{\bf l}) ;q)  \,.\nonumber\\
\ee Here we use the $R'$ R-symmetry, see Figure \ref{fig:3bbA1}, and $\epsilon({\bf m})=(1-(-1)^{2{\bf m}})/4$. Then the index of a sphere with $s_+$ positive punctures, $s_-$ negative punctures, and flux ${\frak n}$ is given by,
\be
&&{\cal I}_{{\frak n},s_+,s_-}(\{w^+,w^-,F\},\{{\bf n}^+,{\bf n}^-\},G;q,t)=\nonumber\\
&&\sum_{{\bf m}\in  \frac12{\mathbb Z}}\oint\frac{dz}{4\pi i z}(1-q^{|m|} z^{\pm2})q^{-|{\bf m}|}G^{\frac{1-(-1)^2{\bf m}}2}\\
&&\prod_{i=1}^{{\frak n}-\frac12 s_++\frac12 s_-} {\cal I}_{\frac12}(t^{-1}z^{\pm2}F_i,\pm2{\bf m};q){\cal I}_{\frac12}(t^{-1}F_i,0;q)\times\nonumber\\
&&\prod_{j=1}^{s_+}{\cal I}(w^+_j,{\bf n}^+_j|z,{\bf m};q,t^{-1})\prod_{h=1}^{s_-}{\cal I}(w^-_j,{\bf n}^-_j|z,{\bf m};q,t)\,.\nonumber
\ee Here $(w^{\pm},{\bf n}^{\pm})$ are the fugacities and fluxes for the puncture symmetries, fugacities $F_i$ are for $SU({\bf n})$ symmetry, and $G$ is the fugacity for the ${\mathbb Z}_2$ global symmetry. This index can be checked to be equal, at least in expansion in $q$, to the index computed in the dual frame using the building blocks of Figure \ref{fig:3bbA1}. For example for the case of no punctures and ${\bf n}=3$ the index of the dual is,
\be
&&\sum_{m\in {\mathbb Z}} \oint\frac{dz}{4\pi i z}(1-q^{|m|} z^{\pm2})q^{-|{\bf m}|}\times \\
&&{\cal I}_{\frac12}(t^{-1}z^{\pm2},\pm2{\bf m};q){\cal I}_{\frac12}(t^{-1},0;q) \nonumber\\
&& {\cal I}_{\frac34}(t^{\frac12}z^{\pm1} G,\pm{\bf m};q) {\cal I}_{\frac14}(t^{\frac32}z^{\pm1} G,\pm{\bf m};q)  {\cal I}_1(t^{-2} ,0;q)^2\nonumber\\
&& {\cal I}_{\frac34}(t^{\frac12}z^{\pm1} ,\pm{\bf m};q) {\cal I}_{\frac14}(t^{\frac32}z^{\pm1},\pm{\bf m};q)  {\cal I}_1(t^{-2} ,0;q)^2\nonumber
\ee Here the first two lines correspond to to the single gauge node and the last two lines to the two one punctured spheres we glue together. The fugacity $G$ appears only in one of the two 
one punctured spheres. This index can be checked to be equal to ${\cal I}_{{\frak n}=3,0,0}(\{\emptyset,\emptyset,1\},\{\emptyset,\emptyset\},G;q,t)$. Note that we cannot match the $SU({\frak n})$ symmetry as this is emergent on side $A$.

\subsection{Topological index}

Next we consider the topological index, or $\Sigma_g \times S^1$ partition function.  Here we take a topological twist along $\Sigma_g$.  We can use this index as a detailed check of some of the dualities we propose.  In particular, as it is sensitive to global structure of the gauge group, it can detect some of the global issues discussed in Section \ref{sec:global}.

The topologically twisted index is defined and discussed in \cite{Nekrasov:2014xaa,Benini:2015noa,Closset:2016arn}.  Here we recall that for a gauge theory with simply connected gauge group $G$, flavor symmetry $G_F$, and matter in representation $R$ of $G \times G_F$, we introduce fugacities $x_a$, $a=1,...,r_G$, and $\nu_i$, $i=1,...,r_{G_F}$, associated to the Cartan subalgebras of the gauge and flavor groups, and write ``Bethe equations'':\footnote{Here we use the so-called ''$U(1)_{-1/2}$ quantization'' for chiral multiplets which preserves gauge invariance at the expense of introducing a parity-breaking regulator that behaves like a ``level $-\frac{1}{2}$ CS term.''  Below we will implicitly shift the effective CS terms by including appropriate bare CS terms.}
\be \Pi_a \equiv x_b^{k_{ab}} \prod_{(\rho,\omega) \in R} (1-x^\rho \nu^\omega)^{-\rho_a} = 1, \;\;\;a =1,...,r_G \;\;\;\;\;\ee
where $k_{ab}$ is  the matrix of (bare) Chern-Simons levels for $G$.  The solutions, modulo Weyl symmetry, are in one-to-one correspondence with the supersymmetric vacua of the 3d gauge theory on $\R^2 \times S^1$.  Then the $\Sigma_g \times S^1$ partition function is given by a sum over solutions to these equation of certain insertions which add handles and flavor flux on $\Sigma_g$.  

When $G$ is not simply connected, these equations are slightly modified, as discussed in \cite{BWHF}.  This will play an important role when mapping discrete symmetries across the duality, and we discuss this in some simple cases below.

\

\noindent {\bf Sphere with flux $\n$} 

\

Let us consider the theory $\cT_{3d}[{S_\n}^2 \times S^1]$ associated to the sphere with flux $\n \geq 2$.\footnote{We recall the theory with $\n=0,1$ is ill-defined, and the theory with $\n \rightarrow -\n$ can be obtained by a $U(1)_t$ charge conjugation.}  Above we showed that this theory has two dual description, the ``star-shaped quiver'' description, which in this case is simply an $SO(3)$ gauge theory with $\n$ fundamental chiral multiplets, and an $SU(2)^{\n-2}$  linear quiver gauge theory.  Here we compare the partition functions computed using these two descriptions.  

\

{\it Star-shaped quiver}

\

We first consider the star-shaped quiver description  This theory has either $SU(2)$ or $SO(3)$ gauge group, depending on the choice of higher form symmetry structure, as discussed in Section \ref{sec:global}.  In addition, there are $\n$ adjoint chiral multiplets, with charges under $U(1)_R$ and $U(1)_t$ given in Table \ref{tab:ssqcharges}.  Here we note that the R-symmetry, $R'$, of Section \ref{sec:4dmodels} is related to the one obtained by reduction of the 6d $U(1)_R \subset Sp(1)_R$ symmetry by:
\be R' = R + \frac{1}{2} T \ee
However, for the purposes of computing the topological index, we will require all R-charges and flavor charges to be integer quantized, and so we introduce new symmetries:
\be \hat{R} = R' + \frac{1}{2} T = R+T, \;\;\;\; \hat{T} = 2 T \ee
We denote the fugacity for the $U(1)_{\hat{t}}$ symmetry by $\tau$.
\begin{table}
	\begin{centering}
		
		\begin{tabular}{|c||c|c|c||c|c|}
			\hline
			Field & $SO(3)_{gauge}$ & $U(1)_t$ & $U(1)_{R}$ & $U(1)_{\hat{t}}$ & $U(1)_{\hat{R}}$ \\
			\hline
			$\phi_{a=1}^\n$ & $3$ & $-1$ & $1$  & $-2$ &  $0$ \\
			\hline
		\end{tabular}

	\end{centering}
	\caption{Charges for star-shaped quiver \label{tab:ssqcharges}}
\end{table}

Let us first write the Bethe equations in the case where the group is $SU(2)$:
\be \label{su2be}\Pi =  \bigg( \frac{x^2 \tau^2 - 1}{\tau^2- x^2} \bigg)^{2\n} = 1 \ee
Here for simplicity we have not included background gauge fields for the full $U(\n)$ flavor symmetry.  This equation has $4f$ solutions for $x$:
\be x_{a,\pm} = \pm \sqrt{ \frac{1+ \tau^2 \zeta^a}{\tau^2+\zeta^a}} , \;\;\; \zeta = e^{\frac{2 \pi i}{2\n}} , \;\;\; a = 0,...,2\n-1 \nn \ee
However, when $x=\pm 1$, corresponding to $a=0$, the vacuum is lifted by fermion zero modes arising from the gauginos.  Moreover, the solutions with $x \rightarrow x^{-1}$, or $a \rightarrow -a$, are related by Weyl symmetry, and so correspond to a single physical vacuum.  Then we find:
\be \cS^{SU(2)}_{vac}=\{ x_{a,\pm}  , a=1,...,\n-1 \} \cup \{ x_{\n,+} = i \sim x_{\n,-} \} \nn  \ee
where we note the solutions $x_{\n,\pm}=\pm i$ are Weyl equivalent, and so give a single vacuum, for a total of $2\n-1$ vacua. 

Next we consider the $SO(3)$ version of the theory.  The modification of the Bethe equations for non-simply connected groups is described in \cite{BWHF}.  In the present case, we must identify states related by large  $SO(3)$ gauge transformations, which act as $x \rightarrow -x$.  In addition, the Bethe equations are modified to:
\be \label{so3be} \bigg(-\frac{\tau^2 x^2 - 1}{\tau^2 - x^2} \bigg)^{\n} = \chi \ee
where $\chi \in \{\pm 1\}$ is a fugacity for the $\Z_2^{\cJ}$ topological symmetry of the $SO(3)$ theory.  Finally, in the state $x_\n$ above there is an unbroken $\Z_2$ gauge symmetry acting, and this $\Z_2$ gauge theory contributes two physical states, which come with $\Z_2^{\cJ}$ charges $\pm 1$.  To summarize, we have:
\be \label{so3ssqstates} \cS^{SO(3),\chi}_{vac}=\begin{cases}
  \{ x_2,x_4,...,x_{\n-2},x_\n^\pm \}   & \n \; \text{even},\; \chi=1, \\  
   \{ x_1,x_3,...,x_{\n-1} \}   & \n\; \text{even}, \; \chi=-1, \\  
   \{ x_1,x_3,...,x_{\n-2}, x_\n^\pm \}   & \n \; \text{odd}, \; \chi=1 ,\\  
 \{ x_2,x_4,...,x_{\n-1} \}   & \n \; \text{odd}, \; \chi=-1. \end{cases} \;\;\;\; \ee
Here we dropped the subscript $\pm$ on $x_{a,\pm}$, as we now identify these solutions, but included a superscript on $x_\n$ accounting for the two $\Z_2$ gauge theory states, with the superscript giving their $\Z_2^{\cJ}$ charge.

Having understood the vacuum structure of the two theories, let us discuss their partition functions.  These are written in terms of the handle-gluing operator and $U(1)_{\hat{t}}$ flux operator.  Using the charges in Table \ref{tab:ssqcharges}, we find:
\bea & \cH_{SU(2)}  & =(x-x^{-1})^{-2} ( (\tau - \tau^{-1} x^{\pm 2})(\tau-\tau^{-1}))^{\n} H  \nn \\
 & \Pi_\tau &= ((\tau- \tau^{-1} x^{\pm 2})(\tau-\tau^{-1}))^{2\n}   \eea
where the Hessian, $H$, is given by:
\be H = x \frac{d \log \Pi}{dx} = \frac{4\n (\tau^4-1)}{(\tau^2-x^2)(\tau^2 -x^{-2})}   \ee
Plugging in the vacua above, we find
\be \label{hposssq} & \cH_{SU(2)}(x_{a,\pm}) &\ds = \frac{4 \n (\tau^4-1)^3 \zeta^a}{\tau^6 (1-\zeta^a)^2} \bigg( \frac{(\tau^2-1)^3 (\tau^2 + 1)^2 \zeta^a}{\tau^3 (\tau^2 + \zeta^a)(1+\tau^2 \zeta^a)} \bigg)^{\n-2} \nn \\
& \Pi_\tau(x_{a,\pm}) & \ds =  \bigg( \frac{(\tau^2-1)^3 (1+\tau^2)^2 \zeta^a}{\tau^3 (\tau^2+\zeta^a)(1+\tau^2 \zeta^a)}  \bigg)^{2\n}  \hspace{2.3cm}  \ee
For the $SO(3)$ theory, we find the same flux operator, but the handle-gluing operators are related by \cite{BWHF}:
\be \label{Hso3} \cH_{SO(3)} = \left\{ \begin{array}{cc} \frac{1}{4} \cH_{SU(2)} & \text{trivial vacuum} \\ \cH_{SU(2)} & \text{$\Z_2$ gauge theory vacuum} \end{array}\right. \ee

Then the partition function on $\Sigma_g \times S^1$ with $U(1)_{\hat{t}}$ flux $\m$ through $\Sigma_g$ is given by:

\be Z^{\Sigma_g \times S^1}_{SU(2)} =  \sum_{\hat{x} \in \cS_{vac}^{SU(2)}} \cH_{SU(2)}(\hat{x})^{g-1} \Pi_\tau(\hat{x})^{\m} \nn \\ Z^{\Sigma_g \times S^1}_{S0(3),\chi} =  \sum_{\hat{x} \in \cS_{vac}^{SO(3),\chi}} \cH_{SO(3)}(\hat{x})^{g-1} \Pi_\tau(\hat{x})^\m \ee
We will consider some simple examples when we compare to the dual linear $SU(2)$ quiver next.  From the form of the expressions above, it suffices to find a one-to-one map between the vacua of the two theory and check that $\cH$ and $\Pi_\tau$  match in dual vacua, which then implies the topological index matches for all $g$ and $\m$.

\subsubsection*{$SU(2)$ linear quiver}

As discussed in the previous section, this can be described by a linear $SU(2)$ quiver gauge theory with $\n-2$ gauge nodes.  Adjacent nodes are connected by a bifundamental chiral multiplet, and the two final nodes contain two fundamental chirals.  Each node also contains an adjoint chiral multiplet.  Finally, there are several singlet fields and superpotential couplings

\

\noindent{\it $\n=2$ case }

\

In this case, the theory is a Wess-Zumino model, with superpotential given by \eqref{f2sup}.  In addition to the $U(1)_t$ and $U(1)_{R'}$ symmetries, in the case there is an explicit $U(2)$ symmetry preserved, and the charges shown in Table \ref{f2tab}.  In addition, this theory admits a discrete $\Z_2$ symmetry acting on $Q$ and $\frak{s}$, for which we introduce a parameter $\chi\in \{ \pm 1\}$. Then there is a single vacuum, and we can immediately write down the handle-gluing and $U(1)_{\hat{t}}$ flux operators:
\be  \label{n2lqops} {\cH}=  (\tau^2 -  \tau^{-2})^{3}  \;\;\;\; {\Pi_\tau} =  \frac{(\tau^2 - \tau^{-2})^{12}}{(\tau^{-1} -\chi \tau)^{4}(\tau^{-2} - \chi \tau^2)^4} \nn \;.\ee

Let us compare this to the $\n=2$ SSQ theory above.  Here we identify the $\Z_2^\cJ$ symmetry of that theory with the $\Z_2$ symmetry discussed above.  In particular, for $\chi=1$, we see from \eqref{so3ssqstates} that the SSQ has a single solution, with $a=2$.  For $\chi=-1$, there is also a single solution, now with $a=1$.  Plugging in to \eqref{hposssq}, we may compare these to \eqref{n2lqops}, and find agreement.\footnote{More precisely, we find precise matching for the eigenvalues of the handle-gluing and flux operators, but the multiplicity of these eigenvalues does not agree for $\chi=1$, \ie, we find two states on the SSQ side and only one on the WZ side.  This discrepancy may be due to an additional local action for $\Z_2$ background fields, similar to that appearing in \cite{Cordova:2017vab} in a similar context.  We leave this for future investigation.}

\begin{table}

	\begin{centering}
		
		\begin{tabular}{|c||c|c|c||c|c|}
			\hline
			Field  & $U(2)$ & $U(1)_{t}$ & $U(1)_{R'}$ &  $U(1)_{\hat{t}}$ & $U(1)_{\hat{R}}$ \\
			\hline
			$\tilde{Q}$ & $2$ & $1$ & $\nicefrac{1}{2}$ & $2$ & $1$ \\
			$\Phi$ & $3$ & $-2$ & $1$ & $-4$ & $0$  \\
			${\frak s}$ & $1$ & $2$ & $0$  & $4$ & $1$ \\
			\hline
		\end{tabular}
		
	\end{centering}
	\caption{Charges for $\n=2$ linear quiver	\label{f2tab}}
\end{table}

\

\ 

\

\noindent{\it $\n=3$ case }

\

\begin{table}
	
	\begin{centering}
		
		\begin{tabular}{|c||c|c|c||c|c|}
			\hline
			Field & $SU(2)$ & $U(1)_{t}$ & $U(1)_{R'}$ & $U(1)_{\hat{t}}$ &  $U(1)_{\hat{R}}$ \\
			\hline
			$q_{i=1}^2$ & $2$ & $\nicefrac{1}{2}$ & $\nicefrac{3}{4}$ & $1$ & $1$ \\
			$q_{i=3}^4$ & $2$ & $\nicefrac{3}{2}$ & $\nicefrac{1}{4}$ & $3$ & $1$  \\
			$\Phi$ & $3$ & $-1$ & $\nicefrac{1}{2}$ & $-2$ & $0$  \\
			$s_{a=1}^4$ & $1$ & $-2$ & $1$  & $-4$ & $0$ \\
			\hline
		\end{tabular}
		
	\end{centering}
	\caption{Charges for $\n=3$ quiver \label{tab:f3charges}}
\end{table}

\

Next we consider the case $\n=3$, which has a single $SU(2)$ gauge node with four fundamentals, an adjoint, and some singlet fields, and with superpotential given by a special case of \eqref{LQsuperpot}. The charges are written in Table \ref{tab:f3charges}.   Letting $x$ denote the $SU(2)$ fugacity, the Bethe equations are:
\be \label{f3be} & \ds  \Pi = \bigg(\frac{x - \tau}{1-x \tau}\bigg)^2 \bigg(\frac{x - \tau^3}{1-x \tau^3}\bigg)^2\bigg( \frac{x^2-\tau^{-2}}{1-x^2 \tau^{-2}} \bigg)^2 \nn\\
& \ds= \frac{(x-\tau^3)^2(\tau x+1)^2}{(1-\tau^3 x)(x+\tau)^2} = 1  \;.\ee
where in the second line we have canceled some factors between the numerator and denominator.  Because of these cancellations, there are fewer solutions than for a theory with the same matter content but no superpotential constraints.  Specifically, there is a single solution, up to Weyl invariance, at (in terms of $y=x+x^{-1}$):
\be y = -2 (\tau+\tau^{-1}) \;. \ee 

To treat this cancellation carefully, we introduce a regulator, $\epsilon$, which can be thought of as a fugacity for a formal $U(1)_\epsilon$ symmetry which is incompatible with our choice of superpotential.  This modifies the Bethe equations to:
\be \label{f3bereg}  \Pi=   \frac{(x - e^\epsilon \tau)(x-e^{-\epsilon} \tau)}{(1-e^{\epsilon} x \tau)(1-e^{-\epsilon} x \tau)} \bigg(\frac{x - \tau^3}{1-x \tau^3}\bigg)^2\bigg( \frac{x^2-\tau^{-2}}{1-x^2 \tau^{-2}} \bigg)^2 = 1 \;. \nn \ee
To make contact with our desired theory, we must then take the limit $\epsilon \rightarrow 0$.  For small but non-zero $\epsilon$, we find three vacua at:
\be y \in \bigg \{\!-2 (\tau+\tau^{-1}) + O(\epsilon),\\
 \; \; \tau+ \tau^{-1} \pm \frac{i}{\sqrt{3}} \sqrt{\epsilon} (\tau-\tau^{-1}) + O(\epsilon) \bigg\} \;. \nn \ee
We recover the solution above, plus two additional solutions, which approach $y=\tau+\tau^{-1}$ (or $x= \tau^{\pm}$) as we remove the regulator.

Next we compute the handle-gluing and flux operators.  Using the charges in Table \ref{tab:f3charges}, the handle-gluing operator is given by:
\be \label{f3hdef} \cH = \frac{(\tau-\tau^{-1})(\tau -\tau^{-1} x^{\pm 2})(\tau^2-\tau^{-2})^4}{(x-x^{-1})^2} H  \;,\ee
where the Hessian $H$ is given by:
\be H = \frac{d \log \Pi}{du} \nn \ee
The behavior of $\cH$ is smooth near the solution at $y=-2(\tau+\tau^{-1})$, and we find:
\be \cH|_{y\rightarrow -2(\tau+\tau^{-1})} = 3 \frac{(\tau-\tau^{-1})^6 (\tau+\tau^{-1})^5}{\tau^2+1+\tau^{-2}}  \ee
For the other vacua, note that for $x \approx \tau$, we may approximate the Hessian as:
\be H \approx C + \frac{\tau+x}{\tau-x} + \frac{1}{1-e^\epsilon \tau x^{-1}} -  \frac{1}{1-e^\epsilon \tau^{-1} x} \ee
This has a pole as $x$ approaches $\tau$, which competes with a zero in \eqref{f3hdef}.  After taking the $\epsilon \rightarrow 0$ limit carefully, we find a finite result which is the same for the two vacua:
\be \cH_{y \rightarrow \tau+\tau^{-1}}  =3 (\tau-\tau^{-1})^4 (\tau+\tau^{-1})^5 \ee
One can similarly compute the behavior of the $U(1)_{\hat{t}}$ flux operator, and we find:
\be \Pi_\tau  = \begin{cases} \ds \frac{(\tau-\tau^{-1})^{18} (\tau+\tau^{-1})^{12}}{(\tau^2 + 1 + \tau^{-2})^6} & y \rightarrow -2(\tau+\tau^{-1}) \;, \\
\ds	(\tau-\tau^{-1})^6 (\tau+\tau^{-1})^{12} & y \rightarrow \tau+\tau^{-1} \;.
	\end{cases} \; \nn\ee

Let us now compare to the $\n=3$ case of the SSQ.  Taking first $\chi=1$ there, we see there are now three vacua, one ordinary vacuum, with $a=1$, and two $\Z_2^\cJ$-charged vacua.  Plugging in to \eqref{hposssq}, we see that the vacuum at $y=-2 (\tau+\tau^{-1})$ here precisely matches with the trivial vacuum at $a=1$, while the two states approaching $y=\tau+\tau^{-1}$ match with the $\Z_2$ gauge theory contributions at $a=3$.

We can also introduce the refinement by a $\Z_2$ symmetry of this theory, which maps to the $\Z_2^{\cal J}$ topological symmetry on the SSQ side.  This can be taken to act on two of the fundamental chirals, $q_1$ and $q_3$, and so, introducing a fugacity $\chi \in\{\pm 1\}$, the Bethe equations are modified to:
\be \label{f3be} & \ds  \Pi = \frac{(x-\tau)(x- \chi \tau)(x-\tau^3)(x-\chi \tau^3)}{(1-\tau x) (1- \chi \tau x)(1-\tau^3 x)(1-\chi \tau^3 x)}  \bigg( \frac{ \tau^2 x^2 - 1}{\tau^2  - x^2} \bigg)^2 \nn
 \ee
One may carry out the same regularization procedure as above, now taking $\chi=-1$.  This time one finds a regular vacuum at $y=0$, and two degenerate vacua near $y=-(\tau + \tau^{-1})$.  However, rather than having smooth behavior as we remove the regulator, the flux and handle-gluing operators now diverge in the degenerate vacuum.  We take this as an indication that these vacua do not contribute, and so there is only one vacuum for $\chi=-1$.  Comparing to \eqref{so3ssqstates}, we see this agrees with the SSQ side, and taking the vacuum with $a=2$ there, we find precise agreement for the handle-gluing and flux operators.  

We emphasize that the topological index is sensitive to the global form of the gauge group, and in all cases above it was crucial that we took the SSQ gauge group to be $SO(3)$, rather than $SU(2)$, both for the number of vacua to agree, and for the handle-gluing operators to precisely match, where we used the prescription of \eqref{Hso3}.

\

\noindent {\bf Sphere with one puncture and $\n=\frac{3}{2}$}

\

\begin{table}
	
	\begin{centering}
		
		\begin{tabular}{|c||c|c||c|}
			\hline
			Field & $SU(2)_\mu$ & $U(1)_{t}$ & $U(1)_{\hat{t}}$ \\
			\hline
			$q_1$ & $2$  & $\nicefrac{1}{2}$ & $1$ \\
			$q_2$ & $2$  & $\nicefrac{3}{2}$ & $3$ \\
			$s_{a=1}^2$ & $1$ & $-1$ & $-2$ \\
			\hline
		\end{tabular}
		
	\end{centering}
	\caption{Charges for WZ describing sphere with $\n=\nicefrac{3}{2}$ and one puncture \label{tab:f32WZcharges}}
\end{table}

For an example involving punctures, let us consider the compactification on a sphere  with flux $\n=\frac{3}{2}$ and one puncture.  The 4d model has a description as a WZ model, shown in Figure \ref{fig:3bbA1}, and the fields and  charges are shown in Table \ref{tab:f32WZcharges}.  This theory has a single vacuum.  Let us consider the flavor flux operators for the $U(1)_{\hat{t}}$ and $SU(2)_F$ symmetries, for which we use fugacities $\tau$ and $\mu$, respectively.  In this case, the flux operators are given by
\bea \label{sf32s1flux} \Pi_\mu & \ds =  \frac{(\mu-\tau^3)(\mu-\tau)}{(1-\mu \tau^3)(1-\mu \tau)} \;, \nn \\
 \Pi_\tau & \ds =  \frac{(\tau^4-1)^8}{(1-\mu \tau^3)(1-\mu^{-1} \tau^3)(1-\mu \tau)(1-\mu^{-1} \tau)}  \;. \eea

\begin{table}
	
	\begin{centering}
		
		\begin{tabular}{|c||c|c|c|c||c|}
			\hline
			Field & $SU(2)_1$ & $SU(2)_2$  & $SU(2)_\mu$ & $U(1)_{t}$ & $U(1)_{\hat{t}}$ \\
			\hline
			$q$ & $2$ & $2$ & $2$ & $-\nicefrac{1}{2}$ & $-1$ \\
			$\Psi$ & $3$ & $1$ & $1$ & $-1$ & $-2$ \\
			$\Theta$ & $1$ & $1$ & $3$ & $1$ & $2$ \\
			\hline
		\end{tabular}
		
	\end{centering}
	\caption{Charges for SSQ for $\n=\nicefrac{3}{2}$ and one puncture \label{tab:f32SSQcharges}}
\end{table}

The dual theory is a star-shaped quiver with one $T[SU(2)]$ leg.  From \eqref{lplm}, we see we may take the central node to have a single adjoint chiral multiplet, $\Psi$, with $U(1)_t$ charge $-1$.  We find it convenient to use the dual description of the $T[SU(2)]$ theory as an $SU(2)_{k=1}$ gauge theory with two flavors, reviewed in Appendix \ref{app:tsun} (see Table \ref{tab:tsu2charges} for the fields and charges), as this makes the $SU(2)$ flavor symmetry manifest in the Bethe equations.  Then the theory has $SU(2)_1 \times SU(2)_2$ gauge symmetry, and the fields and charges are shown in Table \ref{tab:f32SSQcharges}.   The Bethe equations are given by:
\bea \Pi_1 & \ds = {x_1}^{2} \bigg( \frac{{x_1}^2\tau^2- 1}{\tau^2 - {x_1}^2} \bigg)^2  \frac{\tau x_1- x_2^{\pm} \mu^{\pm}}{\tau - x_1 {x_2}^{\pm} {\mu}^{\pm}} \;, \nn \\
\Pi_2 & \ds = {x_2}^{-2}  \frac{\tau x_2- x_1^{\pm} \mu^{\pm}}{\tau- x_2 {x_1}^{\pm} {\mu}^{\pm}} \;. \nn \eea
Recall the central node of the SSQ must be taken as $SO(3)$.  Equivalently, we must gauge a suitable $\Z_2$ $1$-form symmetry, and in the present case, one finds this acts on the solutions as:
\be (x_1,x_2) \rightarrow (-x_1,-x_2) \ee
Then after gauging this symmetry, one finds a single solution, which can be conveniently written in  terms of the Weyl and $\Z_2$ 1-form-invariant quantities $y_\pm \equiv y_1 {y_2}^{\pm}$, where $y_i = x_i+{x_i}^{-1}$, as:
\bea y_+=&\ds  \frac{(1+\tau^2)(\tau-\mu)(\tau \mu - 1) (\tau^3 - \mu +\tau^4 \mu + \tau^3 \mu^2)}{\mu(\tau^3-\mu)( \tau^3 \mu-1)}, \nn \\
  y_- & \ds = \frac{\mu (\tau^2+1)}{\tau^3 - \mu + \tau^4 \mu + \tau^3 \mu^2}  \;. \eea
Plugging this solution into the expression for the flux operators, given by:
\bea\Pi_\mu & \ds = {\mu}^{2} \bigg( \frac{{\mu}^2  - \tau}{1- \tau^2 {\mu}^2} \bigg)^2  \frac{\tau \mu- x_1^{\pm} {x_2}^{\pm}}{\tau- \mu {x_1}^{\pm} {x_2}^{\pm}} \;, \nn \\
\Pi_\tau & \ds = \frac{ (\tau^2-{x_1}^{\pm 2})^2}{ (1-\tau^2 {\mu}^{\pm 2})^2} (\tau-  {x_1}^\pm {x_2}^\pm \mu^{\pm} ) \;,\eea
one finds precisely the result in \eqref{sf32s1flux}.

\section{Discussion and Comments}\label{sec:disc}

In this paper we have discussed the compactifications of the $A_1$ $(2,0)$ theory in six dimensions down to three dimensions on a surface with flux times a circle. In particular, the two orders of performing such a reduction give dual theories in three dimensions. There are several ways in which this discussion can be generalized First we can consider the $A_{N-1}$ $(2,0)$ for $N>2$. Here the 6d$\to$5d$\to$3d order of the reduction is completely analogous to what we have done. However, the 6d$\to$4d$\to$3d is more involved as the theories in 4d are, in general, currently lacking a useful description in terms of Lagrangians (however, see  \cite{Gadde:2015xta,Agarwal:2018ejn,Maruyoshi:2016tqk, Razamat:2019vfd} for Lagrangian constructions in some cases). 

 Another, more interesting, venue of generalization is to compactifications of 6d theories with less supersymmetry.  There, at least in some cases,  much has been understood about the compactification from 6d and 4d, and thus it is possible to use this to further compactify to 3d.  However, in these cases the alternative 6d$\to$5d$\to$3d route has not yet been explored in detail. In particular, a very interesting question is whether a useful mirror duality can be derived by following such a route. There are several subtleties with such generalizations, however, which need to be addressed carefully. First, such a route will be most useful if we have an effective 5d gauge theory description. In general, reducing on a circle a 6d SCFT will result in a strongly coupled theory with no known such description, unless one turns on in addition some holonomy for the global symmetry around that circle. Note that for the $(2,0)$ case, no such holonomy was necessary. For example, by turning on appropriate holonomies when compactifying on a circle, 6d $(1,0)$ theories residing on M5 branes probing $ADE$ singularities result in $ADE$ shaped ${\cal N}=1$ quiver gauge theories in 5d.  As we want to further reduce the theory on a surface of vanishing size, there can be orders of limits issues, \ie, scaling holonomies for the circle compactification with the radius of the circle and scaling the size of the surface, which should be treated carefully. 
 
Another subtlety, in analogy to the 4d to 2d reductions discussed in \cite{Gadde:2015wta}, involves  understanding the effective theory in 3d, even when the 5d Lagrangian do not involve scaling any holonomies with the radius. For example, considering minimal 6d SCFTs on a circle with a twist for a discrete symmetries sometimes give a 5d gauge theory \cite{Jefferson:2017ahm,Razamat:2018gro}. Compactifications of these 6d models down to 4d are understood and thus are a natural venue to try and extend the analysis of this paper. In the partition function language, some subtleties that can arise concern the fate of the non-zero dynamical flux sectors in the $S^3_b\times \Sigma_g$ compactification, as well as with the role of the 5d Chern-Simons terms. A more detailed understanding of the instanton corrections may also be important for understanding these cases.  We leave this for future research.

Finally, let us mention that an indirect evidence in favor of having a mirror symmetry description can be derived by studying (limits of) the superconformal index of theories obtained by compactifying the 6d theories to four dimensions. For example, taking the $(2,0)$ theory the index of theories obtained in 4d can be written as correlator in a TQFT \cite{Gadde:2009kb} of the Riemann surface with a schematic form of \cite{Gadde:2011uv},
\be
{\cal I} = \sum_\lambda C^{2g-2+s}_\lambda\prod_{j=1}^s \psi_\lambda({\bf a}_j)\,.
\ee Here $g$ is the genus of the surface and $s$ is the number of (maximal) punctures. The sum is over a certain set of parameters $\lambda$; \eg, for the type $A_{N-1}$ $(2,0)$ theory these parameterize finite dimensional irreducible representations of $SU(N)$. The quantity $C_\lambda$ is a certain function of fugacities coupled to the space-time and R-symmetries, while $\psi_\lambda({\bf a}_j)$ also depend on the fugacities ${\bf a}_j$ for the global symmetry associated with $j$th puncture. Studying the 3d limit of this index, which entails taking all the fugacities to $1$, the index becomes an $S^3$ partition function as discussed in \cite{Niarchos:2012ah,Dolan:2011rp,Gadde:2011ia,Aharony:2013dha}. Moreover, very interestingly as was shown in \cite{Nishioka:2011dq} (at least in certain limits of the fugacities), the quantities $\psi_\lambda({\bf a})$ become the $S^3$ partition functions of the legs of the star-shaped quiver whereas the discrete sum with the functions $C_\lambda$ becomes the integration coming from the gauging of the central $SU(N)/{\mathbb Z}_N$ node, including the contribution of the adjoint fields.  This can be viewed as an indirect indication for the existence of the star-shaped quiver mirror dual. Now, the same structure of the (limits) of the index can be derived starting from more general 6d theories, see for example \cite{Gaiotto:2015usa} for the case of M5 branes probing ${\mathbb Z}_k$ singularities. Reducing this to three dimensions will give a ``star-shaped'' structure for the relevant $S^3$  partition function, which would be very interesting to understand in terms of a partition function of a physical theory with a star-shaped structure directly in three dimensions.

\

\noindent{\bf Acknowledgments}:~
We would like to thank  Julius Eckhard, Heeyeon Kim, Anton Nedelin, Evyatar Sabag, Sakura Sch\"afer-Nameki, and Gabi Zafrir for very useful conversations.  We would like also to thank Sara Pasquetti for very useful comments on the draft of the paper.  The research of SSR was supported by Israel Science Foundation under grant no. 2289/18, by I-CORE  Program of the Planning and Budgeting Committee, and by BSF grant no. 2018204.   BW was supported by the Simons Foundation Grant \#488629 (Morrison) and the Simons Collaboration on ``Special Holonomy in Geometry, Analysis and Physics.''
\appendix

\section{Dual description of $T[SU(N)]$}
\label{app:tsun}

\begin{figure}
	\begin{center}
		\includegraphics[scale=0.46]{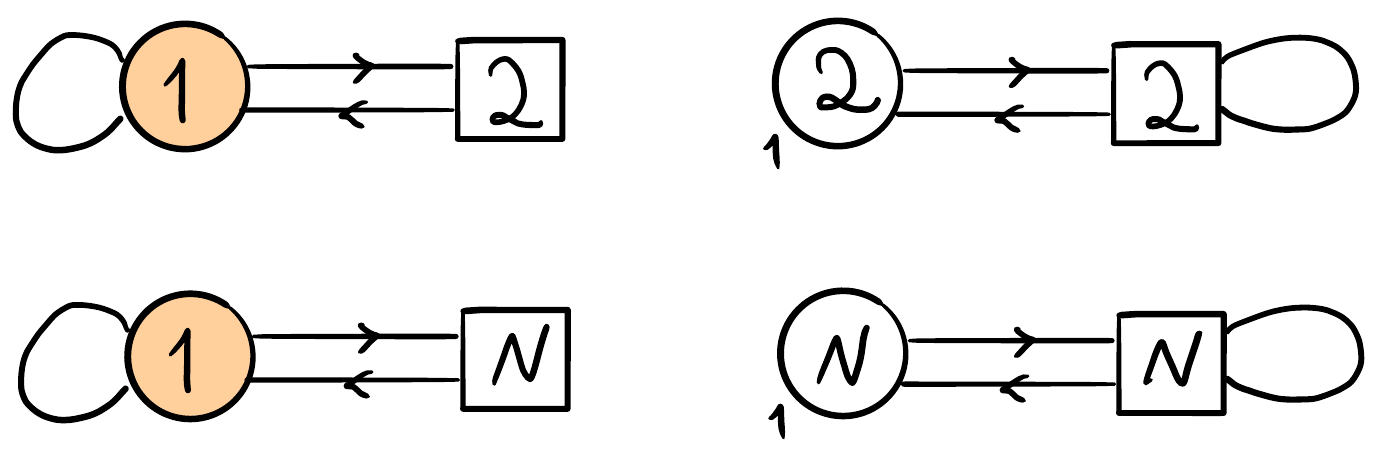}
		\caption{The $T[SU(2)]$ theory. First we have ${\cal N}=4$ description with emergent $SU(2)$ symmetry and then we have an ${\cal N}=2$ description with manifest $SU(2)$. The filled circles represent $U(N)$ groups, while the empty ones represent $SU(N)$. On the bottom there is a generalization to higher rank. Here we gauge $SU(N)$ symmetry with level one of the free three punctured sphere corresponding to sphere with two maximal $SU(N)$ and a minimal $U(1)$ puncture. We have an example of enhanced supersymmetry but not symmetry there.} \label{fig:otherdisc}
	\end{center}
\end{figure}

In this appendix we collect some additional comments about the $T[SU(N)]$ theory.
In addition to the description we discussed in the bulk of the paper, the $T[SU(2)]$ theory has a dual description \cite{Teschner:2012em} which has manifest $SU(2)\times SU(2)$ flavor symmetry but only ${\cal N}=2$ supersymmetry and no manifest time reversal symmetry. The dual is an $SU(2)$ gauge theory with Chern-Simons level one and four fundamental fields coupled through superpotential involving gauge singlet fields. The superpotential breaks the $U(4)$ symmetry to $SU(2)\times SU(2)\times U(1)$, with the two $SU(2)$ symmetries being the Coulomb and Higgs symmetries and the $U(1)$ dual to the Cartan of the enhanced R-symmetry, see Figure \ref{fig:otherdisc}.  The fields and charges of the usual $T[SU(2)]$ description and this dual description are shown in Table \ref{tab:tsu2charges}.

\begin{table}
	
	\begin{centering}
		
		\begin{tabular}{|c||c|c|c||c|}
			\hline
			Field & $SU(2)_H$ & $U(1)_C \subset SU(2)_C$  & $U(1)_t$ & $U(1)_{gauge}$\\
			\hline
			$q$ & $2$ & $0$ & $-\nicefrac{1}{2}$ & $1$ \\
			$\tilde{q}$ & $2$ & $0$ & $-\nicefrac{1}{2}$ & $-1$ \\
			$\Phi$ & $1$ & $0$ & $1$ & $0$ \\
			\hline
		\end{tabular}
	
			\begin{tabular}{|c||c|c|c||c|}
		\hline
		Field & $SU(2)_H$ & $SU(2)_C$  & $U(1)_t$ & $SU(2)_{gauge}$\\
		\hline
		$Q$ & $2$ & $2$ & $-\nicefrac{1}{2}$ & $2$ \\
		$\Theta$ & $1$ & $3$ & $1$ & $1$  \\
		\hline
		CS levels & $-1$ & $-1$ & - & $1$  \\
		\hline
	\end{tabular}
		
	\end{centering}
	\caption{Charges for the two dual descriptions of $T[SU(2)]$.  The top table describes the usual description as $\cN=4$ $U(1)$ with  $N_f=2$, and the bottom describes the $SU(2)_{k=1}$ theory with $N_f=2$, which includes an  additional adjoint ``flip field,'' $\Theta$.  Here we use the same $U(1)_t$ symmetry as in the main text, which is related by a sign to the more standard convention in the literature, and we recall $SU(2)_{H,C}$ are the Higgs and Coulomb flavor symmetries, not to be confused with the R-symmetry. \label{tab:tsu2charges}}
\end{table}

\begin{figure}
	\begin{center}
		\includegraphics[scale=0.48]{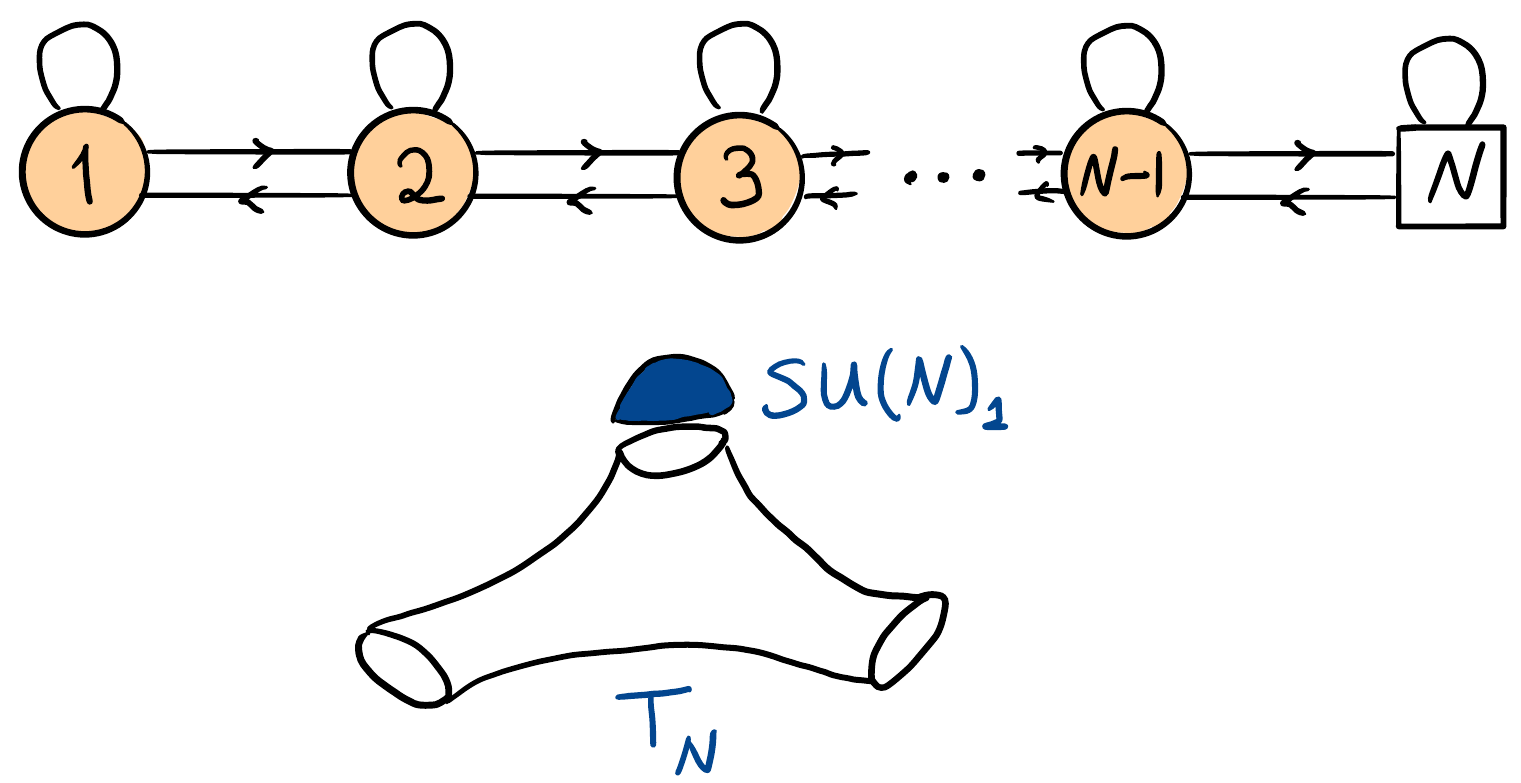} 
		\caption{The construction of the dual of $FT[SU(N)]$ theory by gauging one of the puncture symmetries. On the bottom is the general claim that gauging $SU(N)$ with level one of $T[SU(N)]$ theory one obtains a TQFT which is background Chern-Simons level one theory for the other $SU(N)$.}
	\end{center}
\end{figure}

\begin{figure}
	\begin{center}
		\includegraphics[scale=0.48]{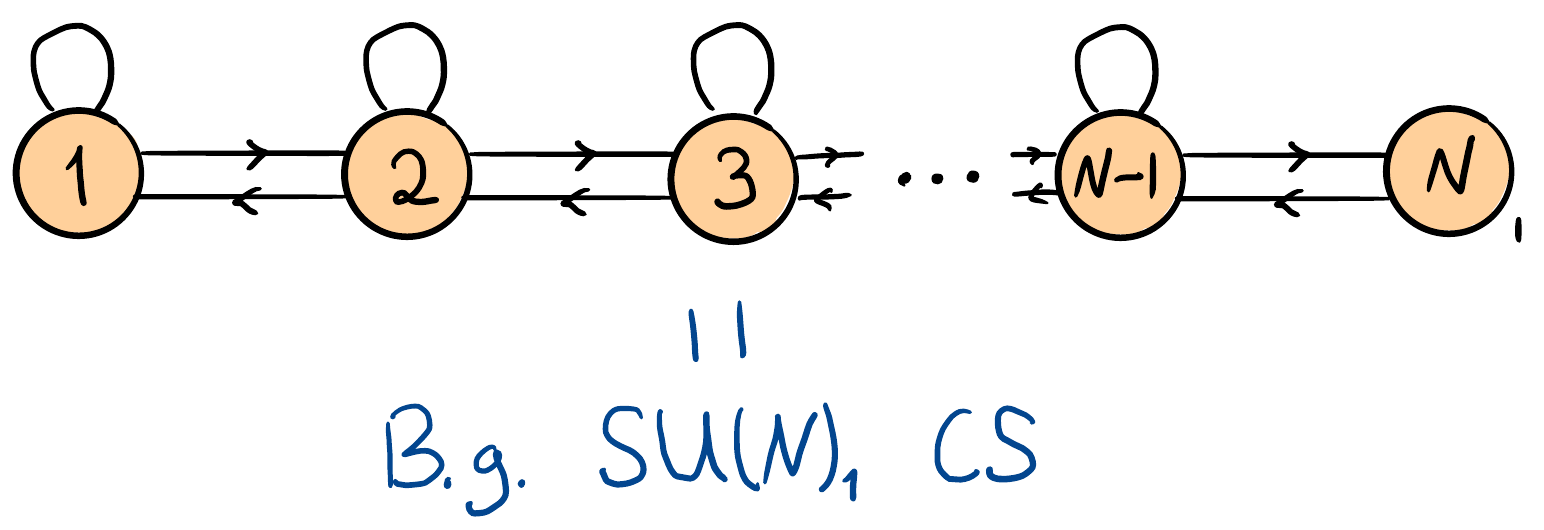} 
		\caption{Gauging $SU(N)$ with level one of $T[SU(N)]$ theory one obtains a TQFT which is background Chern-Simons level one theory for the other $SU(N)$. The theory also contains a decoupled TQFT sector  which is not visible in index computations but can be detected by computing $S^3$ partition functions, see {\it e.g.} \cite{Jafferis:2011ns},  and topological indices \cite{EKSW}. \label{fig:9}}
	\end{center}
\end{figure}

\begin{figure}
	\begin{center}
		\includegraphics[scale=0.48]{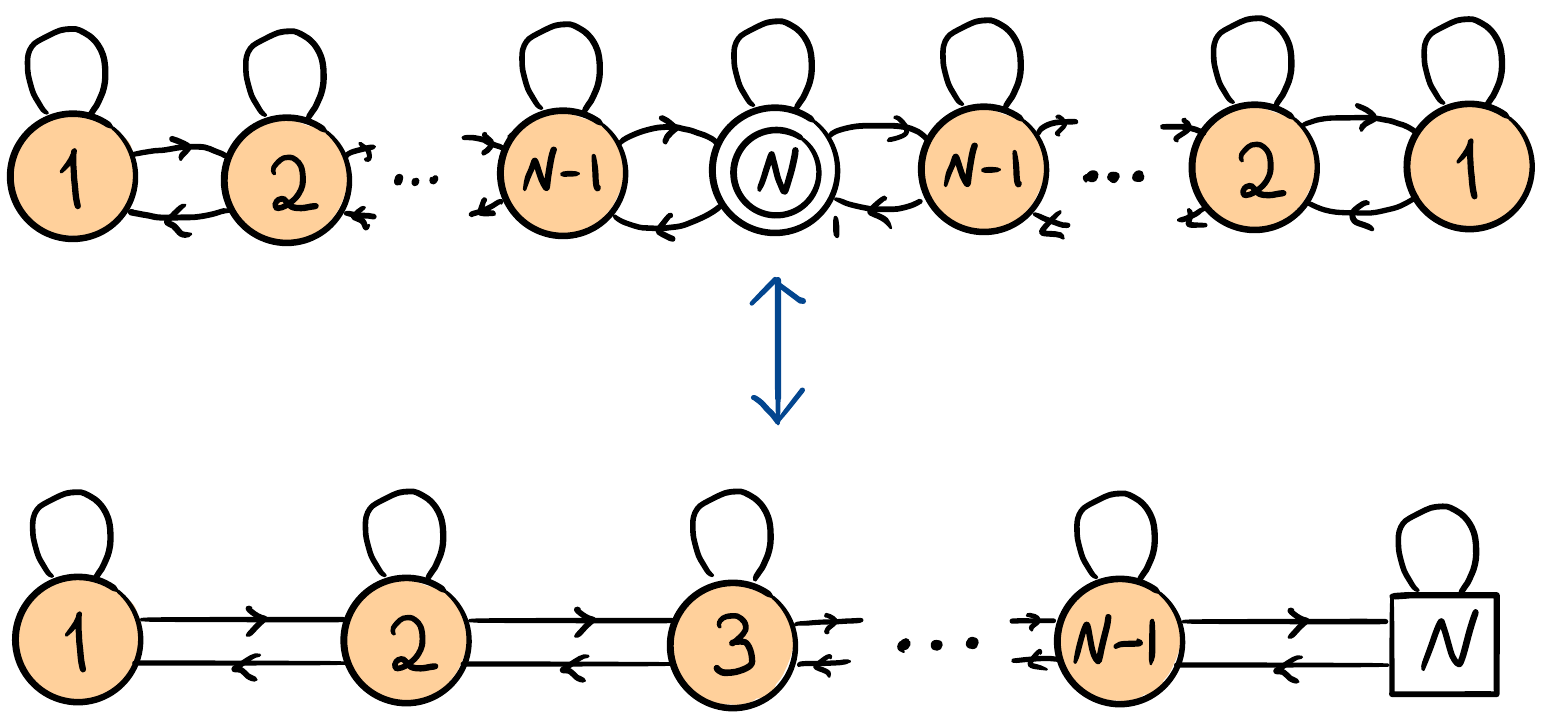}
		\caption{The $FT[SU(N)]$ theory description using star shaped quiver. \label{fig:10}}
	\end{center}
\end{figure}

One can think of the ${\cal N}=2$ description as follows. We can start from the three punctured sphere, which is given by a tri-fundamental of three $SU(2)$s associated with the punctures, and gauge one of the $SU(2)$ symmetries with level one Chern-Simons term. It is well known \cite{Benini:2010uu} that the trifundamental theory is dual to a star-shaped quiver with three copies of $T[SU(2)]$ combined together by gauging a diagonal $SU(2)/{\mathbb Z}_2$ symmetry with ${\cal N}=4$ gauging. Gauging $SU(2)$ with level one Chern-Simons term then in the dual theory corresponds to gauging this symmetry for one of the $T[SU(2)]$s.  It is easier to perform this gauging in the mirror of $T[SU(2)]$ which has manifest $SU(2)$ flavor symmetry. Using a known $USp(2)$ duality \cite{Aharony:1997gp,Giveon:2008zn,Willett:2011gp} this can be shown to be a topological theory equivalent to a contact term for the other $SU(2)$ with level one, tensored with a decoupled $U(1)_2$ topological CS theory (see also \cite{EKSW}), as in Figure \ref{fig:9}. This statement generalizes to arbitrary $T[SU(N)]$. Then one can check that the star shaped quiver with two $T[SU(2)]$ legs with level one Chern-Simons term for the central node is actually equivalent to $T[SU(2)]$ theory with one of the puncture symmetries flipped, shown in Figure \ref{fig:10}. 
Flipping this symmetry puts the two symmetries of $T[SU(2)]$ on the same footing, meaning the model is invariant under exchanging them \cite{Zenkevich:2017ylb} without acting on other symmetry. Following \cite{Zenkevich:2017ylb}, we denote the theory with one of the $SU(2)$ symmetries coupled to adjoint flip chiral fields as $FT[SU(2)]$.

Thus if one believes the duality between the free tri-fundamental chiral field and the star-shaped quiver one derives the dual description with manifest symmetry. There is also a geometric way to understand this statement following the work of \cite{Gukov:2015sna, Pei:2016rmn}, and this can also be embedded in a more general set of dualities following \cite{EKSW}. The $T[SU(2)]$ theory is the three dimensional model living on the duality wall of ${\cal N}=4$ $SU(2)$ gauge theory. From the perspective of engineering such model from the $(2,0)$ theory this corresponds to taking the six dimensional geometry to be four flat directions and a torus. The complex structure of the torus corresponds to the holomorphic coupling  of the ${\cal N}=4$ gauge theory and as we have a duality wall we can think of the theory as having one of the circles non trivially fibered in one direction of the four dimensional space. Such fibrations were discussed in \cite{Gukov:2015sna} and the corresponding theories were argued to correspond to adding a Chern-Simons level to the star shaped models. In fact introducing Chern-Simons levels when gauging symmetries of $T[SU(N)]$ was argued to correspond \cite{Gaiotto:2008ak}  to $T$ transformation of the duality group and then the above statement is equivalent to demanding $ ST ST = T^{-1} S^{-1}$. This also explains why $T[SU(2)]$ is inherently three dimensional model, as trying to make one of the circles of the four dimensional model not compact will make it small across the domain wall.  

These statements can be generalized to higher rank groups. However, at the moment we do not have a general Lagrangian for three punctured spheres so these do not provide a better description for the $T[SU(N)]$ models.

\end{document}